\documentclass[%
 reprint,
 amsmath,amssymb,
 aps,
 prx,
]{revtex4-2}
\usepackage{xr-hyper}
\usepackage{graphicx}
\usepackage[english]{babel}
\usepackage{svg}
\usepackage{xcolor}
\graphicspath{{Figures/}}
\usepackage{dcolumn}
\usepackage{bm} 
\usepackage{physics}
\usepackage{booktabs}
\usepackage{subcaption}
\usepackage{cprotect}
\usepackage{hyperref}
\hypersetup{colorlinks=true,allcolors=blue}
\usepackage{float}
\usepackage{siunitx}
\usepackage[capitalise]{cleveref}
\newcommand{\citepl}{(}
\newcommand{\citepr}{)}
\crefrangelabelformat{subfigure}{#3#1#4--#5\citepl\crefstripprefix{#1}{#2}#6}
\crefrangelabelformat{equation}{#3\citepl#1#4\citepr--\citepl#5\crefstripprefix{#1}{#2}#6\citepr}

\crefname{section}{Sec.}{Sec.}

\newcommand\mbf[1]{\mathbf{#1}}
\newcommand{\rhofixm}{\rho_\mu^\mathrm{fix}}
\newcommand{\rhofix}{$\rhofixm$} 
\newcommand{\rhovarm}{\rho_\mu^\mathrm{var}}
\newcommand{\rhovar}{$\rhovarm$} 
\newcommand{\absrhovar}{$\abs{\rhovarm}$} 
\def\newfeature{RAD}
\def\noisestrength{noise amplitude}
\def\AC{\texttt{AC\_1}}

\makeatletter
\newcommand*{\addFileDependency}[1]{
\typeout{(#1)}
\@addtofilelist{#1}
\IfFileExists{#1}{}{\typeout{No file #1.}}
}\makeatother

\newcommand*{\addexternaldocument}[1]{%
\externaldocument[][nocite]{#1}%
\addFileDependency{#1.tex}%
\addFileDependency{#1.aux}%
}

\begin{document}
\makeatletter

\makeatother 

\title{Tracking the distance to criticality in systems with unknown noise}
\author{Brendan Harris}
\affiliation{School of Physics, The University of Sydney, NSW 2006, Australia}
\author{Leonardo L. Gollo}%
\affiliation{%
The Turner Institute for Brain and Mental Health, School of Psychological Sciences, and Monash Biomedical Imaging, Monash University, Victoria 3168, Australia
}
\author{Ben D. Fulcher}
\affiliation{
 School of Physics, The University of Sydney, NSW 2006, Australia
}
\date{\today}

\begin{abstract}
Many real-world systems undergo abrupt changes in dynamics as they move across critical points, often with dramatic and irreversible consequences.
Much existing theory on identifying the time-series signatures of nearby critical points---such as increased signal variance and slower timescales---is derived from analytically tractable systems, typically considering the case of fixed, low-amplitude noise.
However, real-world systems are often corrupted by unknown levels of noise that can distort these temporal signatures.
Here we aimed to develop noise-robust indicators of the distance to criticality (DTC) for systems affected by dynamical noise in two cases: when the \noisestrength{} is either fixed, or is unknown and variable across recordings.
We present a highly comparative approach to this problem that compares the ability of over 7000 candidate time-series features to track the DTC in the vicinity of a supercritical Hopf bifurcation.
Our method recapitulates existing theory in the fixed-noise case, highlighting conventional time-series features that accurately track the DTC.
But in the variable-noise setting, where these conventional indicators perform poorly, we highlight new types of high-performing time-series features and show that their success is accomplished by capturing the shape of the invariant density (which depends on both the DTC and the \noisestrength{}) relative to the spread of fast fluctuations (which depends on the \noisestrength{}).
We introduce a new high-performing time-series statistic, the Rescaled Auto-Density (\newfeature{}), that combines these two algorithmic components.
We then use \newfeature{} to provide new evidence that brain regions higher in the visual hierarchy are positioned closer to criticality, supporting existing hypotheses about patterns of brain organization that are not detected using conventional metrics of the DTC.
Our results demonstrate how large-scale algorithmic comparison can yield theoretical insights that can motivate new theory and interpretable algorithms for solving important real-world problems.
\end{abstract}
\maketitle

\addexternaldocument{supp}

\section{Introduction}
\label{sec:Introduction}

A critical point, or bifurcation point, marks the value of some control parameter at which the dynamical properties of a system undergo a qualitative change, such as the appearance or disappearance of an attractor~\cite{Strogatz1994}.
Many phase transitions occur at critical points described by bifurcations in macroscopic models of statistical ensembles or complex systems~\cite{Ma2019a}.
In spin glasses, a net magnetization emerges at a critical value of the temperature order parameter~\cite{Carlson1990}, while in cosmology, the temperature controls the electroweak phase transition and baryon (matter--antimatter) asymmetry \cite{Mazumdar2019}.
The dynamics of complex systems---from the brain to biological swarms---can also be understood in terms of their vicinity to a critical point, with near-critical systems possessing a range of functional advantages~\cite{Munoz:2018dn}.
In the brain, for example, traversing critical points is thought to mediate changes in behavioral and cognitive functions such as memory storage and retrieval, as well as visual attention and processing~\cite{Freyer2012,Cocchi2017,chenComputingModulatingSpontaneous2019,Fontenele2019,Xu2024}.

Crossing a critical point can have dangerous consequences.
For example, generator voltages in power distribution networks undergo bifurcation and transition from fixed-point stability to unsafe oscillation at a critical value of system load~\cite{Ajjarapu1992, Srivastava1995}.
At so-called tipping points, where crossing a critical point leads to instability, the system may also change catastrophically as it rapidly transitions to a new state~\cite{Kuehn2011, Kuehn2013}.
Catastrophes occur in simple physical systems, such as slipping events in wires under stress~\cite{Peters2012}, as well as in models of sleep--wake transitions~\cite{Phillips:2007jh, Yang2016}, transitions to epileptic brain states~\cite{Maturana2020, Liu2023}, and major climate events such as the desertification of North Africa~\cite{Dakos2008}.
Moreover, critical phenomena are studied across scientific domains, including physics~\cite{Gleiser1991, Reatto2007}, neuroscience~\cite{Cocchi2017, Munoz:2018dn}, medicine~\cite{Kianercy2014, Trefois2015}, biology~\cite{Gsell2016}, and engineering~\cite{CotillaSanchez2012}.
The ubiquity of critical phenomena calls for a noise-robust method that can predict how close a system is to a critical point~\cite{Kong2021}: to give warning of epileptic seizure~\cite{Maturana2020, Liu2023}, anticipate power distribution failures~\cite{Ren2015}, forecast imminent climate catastrophes~\cite{Lenton2011}, or to better control cognitive changes using brain stimulation \cite{Lynn:2019}.

Despite the ubiquity and diversity of critical systems, critical phenomena are remarkably linked by a common mathematical foundation: normal forms and bifurcation theory~\cite{Kuehn2011, Kuehn2013}.
This theory categorizes most systems near a critical point into a well-studied, analytical taxonomy of bifurcations~\cite{Strogatz1994, Kuznetsov2004}.
Moreover, a surprising number of systems exhibit relatively simple, codimension-one bifurcations~\cite{Strogatz1994, Zhang2019b}, including saddle-node bifurcations found in the dynamics of neuronal spiking~\cite{Meisel2015}, pitchfork bifurcations in Bose--Einstein condensates~\cite{Rau2010}, or Hopf bifurcations in laser cavities~\cite{Ushakov2005}, financial markets~\cite{Gao2009}, the auditory system~\cite{Hudspeth2014}, thalamocortical models~\cite{Victor2011}, and models of interacting brain regions~\cite{Aquino2022}.
Each type of bifurcation is associated with a normal form---a dynamical equation that encapsulates the essential qualitative behavior near the critical point, regardless of the system~\cite{Crawford1991}.
Driven by the allure of solving a plethora of real-world challenges simultaneously, much recent work has studied this unifying mathematical structure of critical phenomena with the aim of identifying universal indicators of criticality.
For normal forms, and other systems that can be modeled analytically, this task is equivalent to estimating the control parameter that determines how close the system is to the bifurcation point.
As such, a primary goal of studying normal forms has been to derive simple dynamical properties that diagnose how close a critical system is to a bifurcation point.
Here, we term this quantity the \textit{distance to criticality} (DTC) ~\cite{OByrne2022}, which is determined by the control parameter in model systems.

Studies of normal forms and real-world critical systems have identified basic but universal dynamical characteristics that are informative of the DTC~\cite{Scheffer2012}: near a critical point, dynamics are more variable and evolve on a slower timescale~\cite{Scheffer2009a, Gsell2016, Hart2020}.
Both properties can be understood as a byproduct of the flattening of the potential function of a system near the critical point.
A flatter potential about a local minimum (fixed point) increases the return time when a system is perturbed~\cite{Wissel1984} and slows the timescale of fluctuations in a system driven by noise; a phenomenon known as \textit{critical slowing down}~\cite{Scheffer2009a}.
Because a flatter potential function also decreases the confinement of noise-related diffusion, the domain explored by a stochastic system will expand when the system is close to the critical point.
Hence, the standard deviation of a time series increases near a bifurcation point, and has been shown to scale like a negative power of the control parameter when the strength of additive noise is close to zero~\cite{Kuehn2011, Negahbani2015, ORegan2018}.
Although properties of the potential function naturally motivate standard deviation and autocorrelation as indicators of criticality, many related features are used to estimate the DTC in a variety of fields~\cite{Scheffer2012, Dakos2012}, including: time-series skewness of chlorophyll in lake ecosystems~\cite{Carpenter2011}; increases in low-frequency fluctuations of voltage in electrical power systems~\cite{CotillaSanchez2012}; spatial autocorrelation of coupled dynamical systems~\cite{Dakos2009}; and other multivariate critical indicators~\cite{Weinans2021, Liu2023I, Liu2023}.
Notably, \textit{qualitative} signatures of criticality, such as maximized autocorrelation timescales, are universal across classes of bifurcations and forms of noise.

In deriving \textit{quantitative} statistical indicators for the DTC, mathematical analyses typically assume the presence of a weak stochastic component relative to dominant deterministic dynamics~\cite{Arnold1998}.
Stochastic influences are typically classified as either: \textit{measurement noise}, which is applied to the final signal independently from the underlying dynamics of the system; or \textit{dynamical noise}, which is incorporated into the equations of motion and constantly perturbs the deterministic dynamics~\cite{Siefert2003, Sase2016}.
Progress has been made in deriving scaling laws for critical indicators when the strength of dynamical noise is close to zero, such as the relationship between variance and the control parameter when approaching various bifurcations~\cite{Kuehn2011, Kuehn2013}.
Unfortunately, even noise with a small but non-vanishing strength complicates the task of defining (let alone predicting) bifurcation points in stochastic systems~\cite{Crauel1998, Callaway2017}.
One common approach, which we operate under in this work, is to designate the bifurcation point as the value of the control parameter that divides invariant densities with distinct qualitative properties (see \cref{sec:invariantdensity}).
For example, if dynamical noise is incorporated into the normal form of a pitchfork bifurcation, then the distribution of states visited by the system converges over time, for a fixed control parameter.
The transition between a unimodal and bimodal invariant density, corresponding to appearance of a second attractor, occurs at the same value of the control parameter as bifurcation in the deterministic terms alone~\cite{Meunier1988}.
This splitting of the invariant density (see \cref{sec:invariantdensity}), termed a \textit{phenomenological} bifurcation~\cite{Arnold1998}, tends to precede other types of bifurcation points defined by changes in dynamical measures (known as \textit{dynamical} bifurcations)~\cite{Meunier1988}, and gives a natural definition for the critical point that we use here.

Many physical systems, however, are subject to noise that is not small relative to the scale of their deterministic dynamics.
In real-world applications, the limitations of conventional indicators of nearby critical points are tied to the need for quantitative accuracy despite a limited understanding of the system, in particular its stochastic components.
Rather than looking for a qualitative increase or divergence in a statistical indicator of the DTC, which may be sufficient to diagnose a bifurcation after it has taken place, tasks such as estimating the amount of time (e.g., minutes or seconds) until a patient will have an epileptic seizure, or the amount of time (e.g., years) until an ecosystem will collapse, require precisely calibrating the values of a critical indicator against a control parameter.
In such cases, variations in the strength of noise destroy the precise relationships required for quantitatively estimating the DTC, such as by shifting the control parameter at which dynamical measures peak~\cite{Meunier1988}.
For example, moderate noise has been incorporated into models of critical transitions to neuron spiking~\cite{Meisel2015}, climate tipping~\cite{Thompson2011}, and the sleep--wake transition~\cite{Yang2016}.
Although conventional critical indicators are analytically universal (close to the critical point, when noise is vanishingly small) and have been successfully applied to some real-world problems~\cite{Scheffer2012,Dakos2008,Maturana2020}, they have yet to find widespread practical application in many noisy scenarios, such as predicting epileptic seizures~\cite{Wilkat2019,Liu2023}.
Since properties such as standard deviation, autocorrelation, or skewness are highly sensitive to noise~\cite{ORegan2018}, the need for exact calibration is a point of failure for conventional indicators of criticality in systems where the strength of noise is unknown or variable.
Dynamical noise---particularly when it has an unknown variance or distribution---is one of the major obstacles to applying universal theory and critical indicators derived for analytic systems to real-world scenarios involving finite, noisy time series.

To our knowledge, all conventional indicators of criticality are sensitive to noise.
Some prior work has characterized how the strength of a noise process affects the time-series properties used to infer the control parameter, and therefore the DTC.
Although scaling laws, which describe how the indicator varies with the control parameter, have been derived analytically for some indicators of criticality in low-noise cases, such scaling may not hold under moderate or high levels of noise.
For instance, \citet{Meunier1988} demonstrated for the pitchfork bifurcation that features such as the Lyapunov exponent and autocorrelation no longer peak at the phenomenological bifurcation point in presence of additive noise.
Instead, the peak of these time-series features (which marks a dynamical bifurcation point) is shifted along the control parameter to a degree that depends on the noise variance, thereby obscuring the critical point.
Moreover, the splitting of the invariant density becomes less abrupt with increased noise, and other dynamical effects can occur after bifurcation of the underlying deterministic component~\cite{Meunier1988,Fronzoni1987}.
Even a recent measure, formulated to reliably signal the critical point across classes of bifurcations, is sensitive to dynamical noise~\cite{Grziwotz2023}.

While some studies have examined the behavior of critical indicators in systems with a given fixed noise level, prior work has yet to address the arguably more common real-world setting in which the \noisestrength{} is unknown and may vary (across different recordings of the system, or over instances of otherwise similar systems).
For example, variable dynamical noise may arise when studying: (i) the ecological dynamics of species, with noise varying between sites with different climates or weather patterns; (ii) the response of brain areas to random stimulation, where the strength of the stimulation is metered by anatomical differences such as skull thickness~\cite{Groen2019}; (iii) the sleep-wake or epilepsy transition, in which dynamical noise may be influenced by a subject's individual physiology or environment~\cite{Yang2016,Yang2017}; (iv) climate systems, in which the influence of sources of variability such as human activity change over time~\cite{Thompson2010,Thompson2011}; or (v) components of other systems that may be influenced by different external stochastic drives or complex internal heterogeneity~\cite{Mejias2014,Fontenele2019}. %
This is a particularly challenging problem since knowledge of the noise process, especially its variance, is required to properly calibrate time-series features for meaningful prediction of the DTC.
Additionally, established analytical tools that have been used to make progress in idealized cases cannot be applied straightforwardly to the variable-noise setting because their precise relationship to the control parameter depends on the \noisestrength{}.
For example, two prototypical time-series indicators of the DTC---standard deviation and autocorrelation---both depend strongly on the strength of noise in the system: standard deviation increases with noise level, while autocorrelation decreases.

Given the ubiquity of systems corrupted by an uncertain amount of noise, it is of great importance to formulate robust indicators of the distance of such systems to nearby critical points.
In this work, we introduce a data-driven methodology to tackle this problem, in which we simulate a noisy bifurcating system close to a critical point and then search across a large candidate library of time-series features, scoring each on how well it captures the DTC of a simulated system (\cref{sec:FeatureScoring}).
The most comprehensive collection of time-series features to date is the highly comparative time-series analysis library, \textit{hctsa}, which contains over $7000$ diverse features~\cite{Fulcher2013, Fulcher2017, Fulcher:2018fl}.
It contains measures of the distribution of time-series values, self-correlation properties, measures of predictability and complexity, methods related to self-similarity and recurrence properties derived from the literature on physics and nonlinear time-series analysis, among many others.
Here, we address the challenge of inferring the DTC using the data-driven approach of identifying the most informative time-series features for a given criticality problem by searching across the \textit{hctsa} feature set.
To study how each time-series feature is influenced by dynamical noise, we consider our model system in two scenarios (\cref{sec:FeatureScoring}):
(i) the \textit{fixed-noise} case in which the noise has a fixed strength; and
(ii) the \textit{variable-noise} case in which the \noisestrength{} varies across repeated measurements of the system.
While ensemble methods that use combinations of time-series features to optimize a performance metric have been employed in many prior applications of feature libraries such as \textit{hctsa}~\cite{Fulcher2013,Tan2021}, we aimed in this work to demonstrate an alternative approach: that a large library of time-series analysis algorithms can form the foundation for building novel time-series theory and interpretable algorithms for analyzing real-world data.
We find the most noise-robust features by identifying those that covary most strongly with the control parameter in the case of variable noise, then use close inspection to explain how these high-performing features combine the characteristics of noise-driven fast fluctuations with measurements taken from the stationary distribution (see \cref{sec:UnderstandingTopFeatures}).
This combination enables a precise estimate of the shape of the potential---and hence an ability to track the DTC---despite the confound of uncertain noise amplitude.

The paper is structured as follows.
In \cref{sec:Method}, we introduce the supercritical Hopf normal form we use as a model critical system (in \cref{sec:ModelSystem}) and describe how we sampled noise and control parameters to generate a dataset of simulated time series (in \cref{sec:TimeSeries}).
We detail our feature-extraction procedure and our scoring method for identifying high-performing features in the fixed-noise and variable-noise cases in \cref{sec:FeatureScoring}.
Next, in \cref{sec:Results}, we reveal how well features across the \textit{hctsa} library perform in the fixed-noise (\cref{sec:FixedNoise}) and variable-noise (\cref{sec:variablenoise}) cases.
After identifying the top-performing \textit{hctsa} features in the variable-noise case and examining them in detail (\cref{sec:UnderstandingTopFeatures}), in \cref{sec:NewFeature} we summarize the theoretical insight we gained from our data-driven approach and encapsulate our findings in a new time-series feature, which we call \newfeature{}, that can robustly track the DTC in near-critical systems with unknown \noisestrength{}.
In \cref{sec:casestudy} we use \newfeature{} to examine mouse electrophysiology data sourced from the Allen Neuropixels Visual Behavior dataset.
Our investigation tests a structure--function hypothesis positing that regions higher in the cortical hierarchy exhibit longer timescales~\cite{Kiebel2008,Murray2014,Gollo2015,Chaudhuri2015,raut2020} due to their closer proximity to criticality~\cite{Cocchi2017,liang2024}.
Our findings demonstrate the ability of \newfeature{} to accurately track the anatomical hierarchy of visual cortical regions, outperforming conventional indicators of the DTC.
Finally, in \cref{sec:Discussion} we provide concluding remarks, discuss the implications of our findings for other noisy critical phenomena, and outline potential directions for future research.

\begin{figure*}[hp!]
	\centering
 \includegraphics[width=0.98\textwidth]{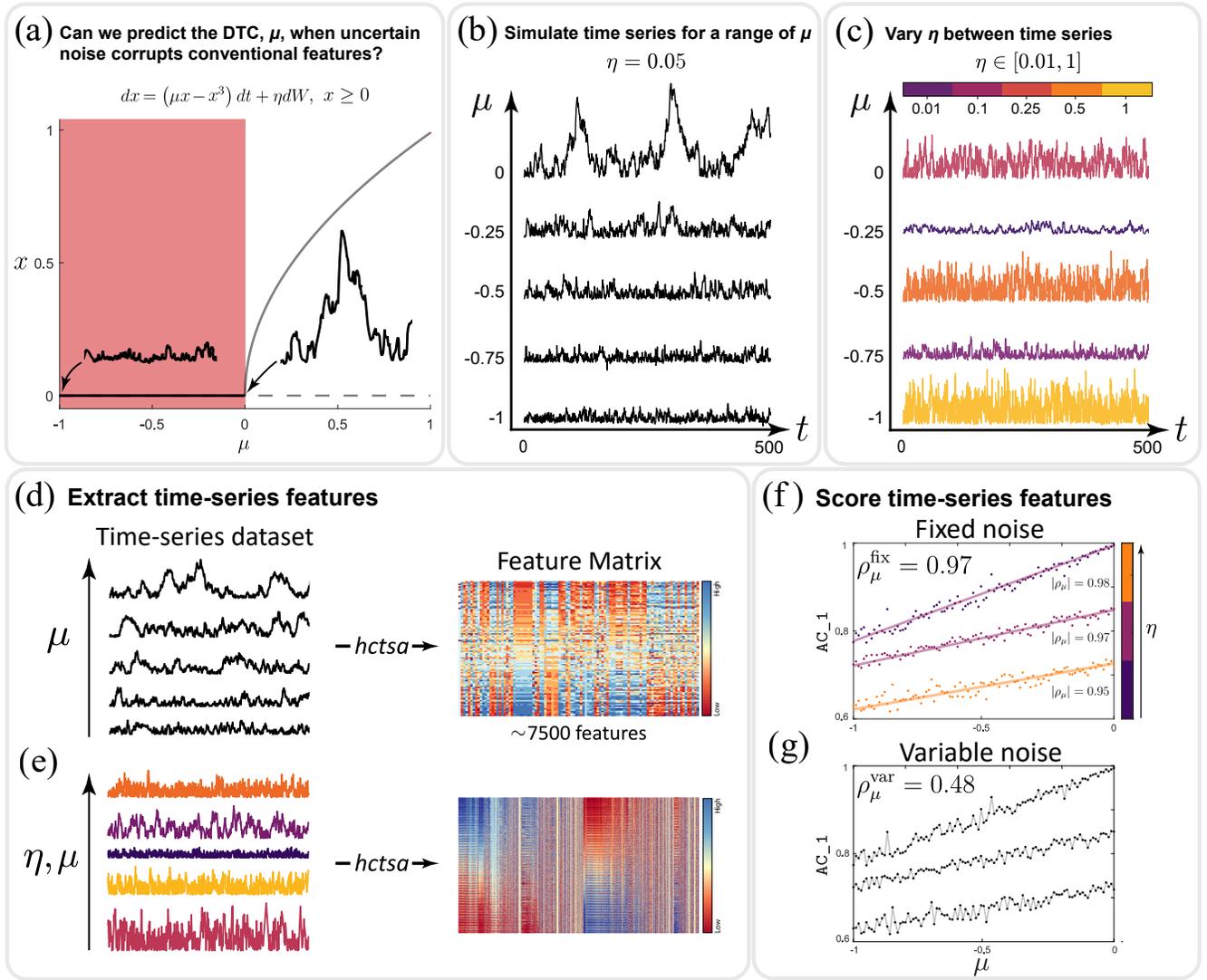}
 {\phantomsubcaption\label{fig:Schematic_a}}
{\phantomsubcaption\label{fig:Schematic_b}}
{\phantomsubcaption\label{fig:Schematic_c}}
{\phantomsubcaption\label{fig:Schematic_d}}
{\phantomsubcaption\label{fig:Schematic_e}}
{\phantomsubcaption\label{fig:Schematic_f}}
{\phantomsubcaption\label{fig:Schematic_g}}
	\cprotect
    \caption{We take a data-driven approach to identifying time-series features that accurately track the distance to a critical point (DTC) of a noisy, near-critical system across a range of noise levels.
	(a) We investigate a model system close to the critical point: the radial component of the normal form for a supercritical Hopf bifurcation with $\mu < 0$, given by \cref{eqn:SupercriticalHopfRadial}.
	A bifurcation diagram shows the radius of equilibrium points $x$, against the control parameter $\mu$---when $\mu < 0$ (shaded red), there is a stable fixed point (solid black) at the origin.
	The model incorporates additive white Gaussian noise, given by $dW$, with a \noisestrength{} of $\eta$.
	We simulated time series across $-1 \le \mu \le 0$ and $0 < \eta \le 1$. Snippets of representative time series are annotated.
	(b) Representative time series of 5000 samples for the fixed-noise problem, for which the noise level is fixed (shown here for $\eta = 0.01$), at selected values of $\mu = -1, -0.75, ..., 0$.
	As $\mu$ approaches zero, it is visually clear that signal variance increases and fluctuations are slower.
	(c) Sample time series for the variable-noise problem: a different value of $0.01 \leq \eta \leq 1$ is chosen for each time series, as indicated by the coloration.
	Relative to the fixed-noise case, the variable \noisestrength{} obscures the underlying variation in the control parameter $\mu$.
	(d, e) To find time-series properties that are sensitive to $\mu$ in both the fixed-noise and variable-noise settings, we compared $7492$ candidate time-series features using \textit{hctsa}~\cite{Fulcher2017}.
	(d) The result of this feature extraction for a fixed value of $\eta$ is depicted as a time series $\times$ feature matrix, with rows ordered by $\mu$.
	(e) The feature matrix for our full dataset, with variation in both $\mu$ and $\eta$, is depicted in the right panel, with rows ordered by $\eta$.
	(f, g) Finally, we scored each feature for its performance under the two conditions as \rhofix{} and \rhovar{}.
	In the fixed-noise setting (f), \rhofix{} was computed by averaging the magnitudes of the correlation of a feature with $\mu$ for each separate value of $\eta = 0.01, 0.02, \dots, 1$ (colored lines). In the variable-noise case (g), \rhovar{}, was an overall correlation to $\mu$ after $\eta$ labels were discarded (indicated in gray).
	An example is shown here for the autocorrelation at lag 1 (labeled \verb|AC_1|), which is strongly correlated to $\mu$ for a given noise-level $\eta$ (f) but only weakly correlated in the variable-noise case (g).
	}
	\label{fig:Schematic}
\end{figure*}

\section{Methods}
\label{sec:Method}

Here we outline our data-driven approach to finding useful statistical indicators of the DTC in the presence of dynamical noise in two cases: the \textit{fixed-noise} case; and the \textit{variable-noise} case.
Our approach is shown schematically in \cref{fig:Schematic}, and can be summarized in two main steps:
(i) simulate a time-series dataset from the normal form of a supercritical Hopf bifurcation [see \cref{eqn:SupercriticalHopfRadial}, below], where each time series is generated using a specific control parameter (varying over a range up to the bifurcation point) and noise variance (either fixed, or varying over a specified range); and
(ii) extract candidate features from each time series using the \textit{hctsa} time-series feature library \cite{Fulcher2013, Fulcher2017} and score each feature on how well it tracks the control parameter $\mu$ across the dataset.
In the fixed-noise case, the noise process has a constant strength, i.e., $\eta$ is fixed for all time series in the dataset.
By contrast, in the variable-noise case the \noisestrength{} can take a different value for any given time series; this setting models real-world situations in which the strength of noise might vary across different recordings of a system, or across different instances of similar systems.

\subsection{The model system}
\label{sec:ModelSystem}

We first describe the model system used for generating time series at various distances, in the control parameter, from a critical point.
We chose the radial component of the normal form for a supercritical Hopf bifurcation with dynamical noise, as it applies to a broad range of real-world systems, including the wake-sleep transition~\cite{Yang2016}, seizure dynamics~\cite{Yang2017}, auditory hair cells~\cite{Ospeck2001,OMaoileidigh2012}, financial markets~\cite{Gao2009}, and many others~\cite{Marsden1976,Freyer2012,Hudspeth2014,Cocchi2017,Deco2017}.
Our system derives from the general normal form of a supercritical Hopf bifurcation~\cite{Schumaker1987}, in which a stable fixed point bifurcates to a stable limit cycle and an unstable fixed point.
Since noise often acts radially with a strength that is independent of the system's phase, we chose to use the simplified radial component of the Hopf normal form, given by
\begin{equation}
\label{eqn:SupercriticalHopfRadial}
dx = \left(\mu x - x^3\right) dt + \eta\, dW\,,\quad x \ge 0\,,
\end{equation}
where $W$ is a Wiener process.
Notably, this resembles the normal form of pitchfork bifurcation with a reflecting boundary at the origin~\cite{Freyer2012}, permitting our results to generalize via a simple transformation to systems that exhibit pitchfork-type bifurcations (such as spin-glass systems~\cite{Carlson1990} or models of epigenesis~\cite{Ferrell2012}).
As shown in the bifurcation diagram in \cref{fig:Schematic_a}, a bifurcation occurs at the critical value, $\mu = 0$:
for $\mu < 0$ there is a single stable fixed point at $x = 0$, whereas for $\mu > 0$ the origin is unstable.
The corresponding potential function of this system, detailed in \cref{sec:Potential}, is unimodal for $\mu < 0$, flattens as $\mu$ is increased to 0, and is bimodal for $\mu > 0$.

Here we consider the more complex problem of approaching the critical point from $\mu < 0$, corresponding to the regime with a single, stable fixed point where the system hovers near the origin [shaded in \cref{fig:Schematic_a}].
We focus our study on this regime for three reasons.
First, during events such as power-system failure~\cite{Ajjarapu1992, Srivastava1995} and epileptic seizures~\cite{Yang2017} the fixed-point regime corresponds to safety, with the presence of low-power fluctuations, in contrast to the dangerous high-power oscillations that occur for $\mu > 0$.
Second, the DTC is much more straightforward to estimate for $\mu > 0$: features that measure simple properties of the distribution, such as the mean or median, are insensitive to $\eta$ when the \noisestrength{} is not large compared to the equilibrium radius [given by $\sqrt{\mu}$ for \cref{eqn:SupercriticalHopfRadial}].
Third, when the noise strength is large enough to mask changes in the distribution caused by varying $\mu > 0$, the system is already sufficiently close to the bifurcation point for the disappearance of the unstable branch to be insignificant.

\subsection{Time-series simulation}
\label{sec:TimeSeries}

To evaluate the performance of individual time-series features at tracking $\mu$ in the presence of a stochastic component with strength $\eta$, we simulated \cref{eqn:SupercriticalHopfRadial} in the range $-1 \le \mu \le 0$.
We studied both the fixed-noise and variable-noise scenarios with a combined time-series dataset generated by varying parameters across ranges $-1 \le \mu \le 0$ and $0 < \eta \le 1$.
In total, we simulated $10\,100$ time series, $\mathbf{x}$, using all combinations of 101 equally spaced values for $\mu = -1, -0.99, \dots, 0$, and 100 equally spaced values of $\eta = 0.01, 0.02, \dots, 1$.
Time series were simulated using the Euler--Maruyama method~\cite{Kloeden1992} over 1000\,s with a time-step of $10^{-3}$\,s.
To avoid the effects of transient dynamics (which are sensitive to the initial condition), we set an initial radius of $x_0 = 0$, coinciding with the stable fixed point of the system. %
Moreover, we discarded the first 500\,s of the integration period, corresponding to an interval far wider than the longest timescale in any of our simulations (estimated using the first zero crossing of the autocorrelation function). %
Finally, we down-sampled the remaining 500\,s to a sampling period of $\Delta t = 0.1$\,s, yielding 5000-sample time series that were analyzed in the remainder of this work.
From this combined dataset, we studied the fixed-noise case by searching for features that vary with $\mu$ when $\eta$ is fixed, as depicted in \cref{fig:Schematic_b}.
In the variable-noise case, depicted in \cref{fig:Schematic_c}, we searched for time-series features that vary with $\mu$ in a way that is consistent over confounding variation of $\eta$.
Finally, note that our data-driven approach is focused on relatively short time series containing 5000 samples, in contrast to other studies that evaluate critical indicators on simulated data~\cite{Grziwotz2023}.
Representative examples of simulated time series are in \cref{fig:Schematic_b,fig:Schematic_c}.

\begin{table*}[ht!]
\caption{Summary of top-performing time-series features from the \textit{hctsa} feature library. The magnitude of \rhofix{} and \rhovar{} indicate how well each feature can track the distance to criticality in the fixed-noise and variable-noise settings, respectively.
Our new features are \texttt{fitSupercriticalHopfRadius\_1} and the rescaled auto-density (\newfeature{}), \texttt{CR\_RAD\_1}.
}
\centering
\addtolength{\tabcolsep}{3pt}
\fontsize{8pt}{8pt}
\begin{tabular}{cccc}
\toprule
  \textbf{Feature name} & \rhofix{} & \rhovar{} & \textbf{Description}\\
  \midrule
  \verb|standard_deviation| & $0.98$ & $0.23$ & Sample standard deviation.\\ 
  \verb|AC_1| & $0.97$ & $0.48$ & Lag-1 autocorrelation.\\
  \verb|DN_RemovePoints_max_01_ac1diff|  & $0.94$ & $-0.88$ & Change in lag-1 autocorrelation from removing the largest 10\% of values.\\
  \verb|SB_MotifTwo_mean_uu| & $0.93$ & $0.88$ & Proportion of pairs of consecutive values that are both above the mean.\\
\verb|PP_Compare_rav2_kscn_olapint| & $0.88$ & $-0.87$ & Change in distribution after two-sample moving-average smoothing.\\
\verb|ST_LocalExtrema_l100_meanrat| & $0.91$ & $-0.90$ & Ratio between the average maximum
and minimum in $100$-sample windows.\\
\midrule
\verb|fitSupercriticalHopfRadius_1| & $0.92$ & $-0.90$ & Potential function curve-fit to a kernel-density estimate of the distribution.\\
RAD & $0.93$ & $0.93$ & Product of the spread of differences and tailedness of the distribution.\\
\bottomrule
\end{tabular}
\label{tab:glossary}
\end{table*}

\subsection{Feature scoring}
\label{sec:FeatureScoring}

We next aimed to determine which types of time-series features, extracted from the noisy time series simulated above, could accurately track variations in the known parameter, $\mu$.
We achieved this in a data-driven way using a comprehensive collection of $7873$ candidate time-series features from the highly comparative time-series analysis toolbox, \textit{hctsa} (version 0.98) \cite{Fulcher2013, Fulcher2017}.
Each time-series feature $f: \mathbb{R}^N \rightarrow \mathbb{R}$ maps an input time series (of $N = 5000$ samples here) to a single, real-valued summary statistic.
The \textit{hctsa} feature set contains methods developed across the interdisciplinary time-series analysis literature, including measures of outliers, periodicity, stationarity, predictability, self-affine scaling, and many others.
We extracted all \textit{hctsa} features from each simulated time series in the dataset described above.
After feature extraction, we removed 381 features from our analysis that were not well-behaved across all time series (e.g., produced \verb|NaN| values or constant outputs across the dataset), leaving 7492 good-quality features.
The result of feature extraction across our full time-series dataset is visualized as a time series $\times$ feature matrix in \cref{fig:Schematic_e}.

Our next goal was to assess the ability of each time-series feature to track the underlying variation of the control parameter $\mu$ across the time-series dataset described in \cref{sec:TimeSeries}.
We scored each feature using a Spearman correlation coefficient $\rho$ between its outputs across the dataset and the corresponding values of $\mu$; features with high $|\rho|$ strongly monotonically track the variation in $\mu$.
As depicted in \cref{fig:Schematic_f,fig:Schematic_g}, we considered the fixed-noise and variable-noise cases separately.

\paragraph{Fixed-noise case}

In the fixed-noise case, where the \noisestrength{} is constant across recordings, we analyzed subsets of time series with the same value of $\eta$.
As depicted in \cref{fig:Schematic_f}, for each set of $100$ time series with a given value of $\eta$, we calculated the Spearman correlation of each feature with $\mu$ (over $\mu = -1, -0.99, \dots, 0$).
We then computed the \textit{fixed-noise score} of a given feature, \rhofix{}, as the mean magnitude of these $100$ Spearman correlation coefficients calculated independently for each noise level.
That is, \rhofix{} measures the average performance of a time-series feature across many settings that each have a unique fixed noise level.

\paragraph{Variable-noise case}

In the variable-noise case, where the \noisestrength{} varies between recordings, we disregarded the $\eta$ labels of each time series (treating them as unknown), as depicted in \cref{fig:Schematic_g}.
We then computed the \textit{variable-noise score} of a given feature, \rhovar{}, as the Spearman correlation between the $10\,100$ feature values (over all time series) and the corresponding values of $\mu$.
In contrast to \cref{fig:Schematic_f}, all points in \cref{fig:Schematic_g} are uncolored, reflecting how the variable noise score is a correlation across all data points, regardless of the noise strength. %
That is, \rhovar{} measures the performance in a single setting where the \noisestrength{} varies across time series.
Unlike the fixed-noise score \rhofix{} (which takes absolute values and is therefore non-negative by construction) the variable-noise score \rhovar{} can be positive or negative, indicating whether the feature increases or decreases with the proximity to criticality.

\section{Results}
\label{sec:Results}

By systematically exploring a supercritical Hopf bifurcation, our main aim was to investigate which types of time-series properties (from a diverse library of over $7000$ candidates in \textit{hctsa} \cite{Fulcher2017}) accurately track the distance to criticality (DTC), given by $\abs{\mu}$, in the pre-critical regime of $\mu < 0$.
As detailed in \cref{sec:FeatureScoring} above, features were scored based on their correlation to $\mu$ using either \rhofix{}, in the fixed-noise setting (\cref{sec:FixedNoise}), and, separately, \rhovar{}, in the variable-noise setting (\cref{sec:variablenoise}).
In the variable-noise case, we find surprising types of features that strongly vary with the DTC.
By examining the algorithmic elements that are common among these features and summarizing the theoretical insight they provide (\cref{sec:UnderstandingTopFeatures}), we develop a new time-series feature for tracking the DTC, which we name the \textit{rescaled auto-density} (\newfeature; see \cref{sec:NewFeature}).

\begin{figure*}[ht!]
	\centering
	\includegraphics[width=\textwidth]{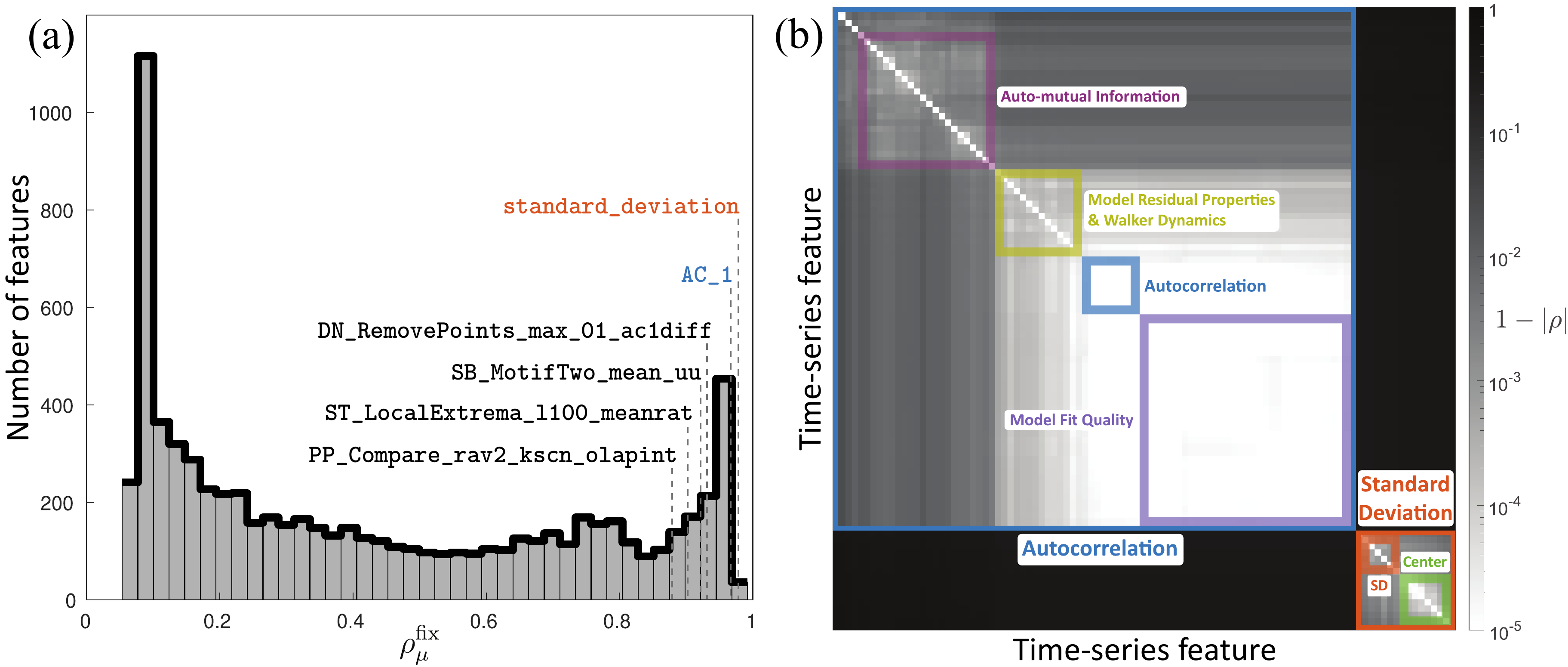}
    {\phantomsubcaption\label{fig:FixedNoiseFeatureHistogram}}
	{\phantomsubcaption\label{fig:featureSimilarityFixed}}
	\cprotect\caption{For a near-critical system with a fixed noise level, we recovered high-performing conventional statistical indicators of the distance to criticality by comparing the behavior of over 7000 candidate time-series features.
	(a) A histogram of fixed-noise feature scores, \rhofix{}, across $7492$ time-series features (from the \textit{hctsa} feature library).
	Selected high-performing features are annotated, including standard deviation (\rhofix{} $= 0.98$, labeled as \verb|standard_deviation|, blue) and lag-1 autocorrelation (\rhofix{} $= 0.97$, \verb|AC_1|, red).
	Names of the top features for the variable-noise case are also annotated for comparison (cf. \cref{fig:VariableNoisePanels}).
	(b) The top 100 features, with \rhofix{} $\ge 0.97$, are plotted as a pairwise correlation matrix.
	The brightness of each element corresponds to the similarity between each pair of these features, using the metric $1 - \abs{\rho}$, where $\rho$ is the Spearman correlation coefficient.
	Rows and columns have been reordered to place similar features close to one another using average hierarchical linkage clustering (on Spearman correlation distances), revealing two clusters of features with common behavior which have been annotated using transparent colored squares.
	As labeled, the first cluster contains features measuring properties of signal autocorrelation (orange), while the second cluster contains features measuring properties related to the signal variance (blue).
    Features annotated in (a) are detailed in \cref{tab:glossary}, and for (b) a sorted list of features is in Supplemental Table~S2.
 	}
	\label{fig:fixedNoisePanels}
\end{figure*}

\begin{figure*}[ht!]
	\centering
	\includegraphics[width=\textwidth]{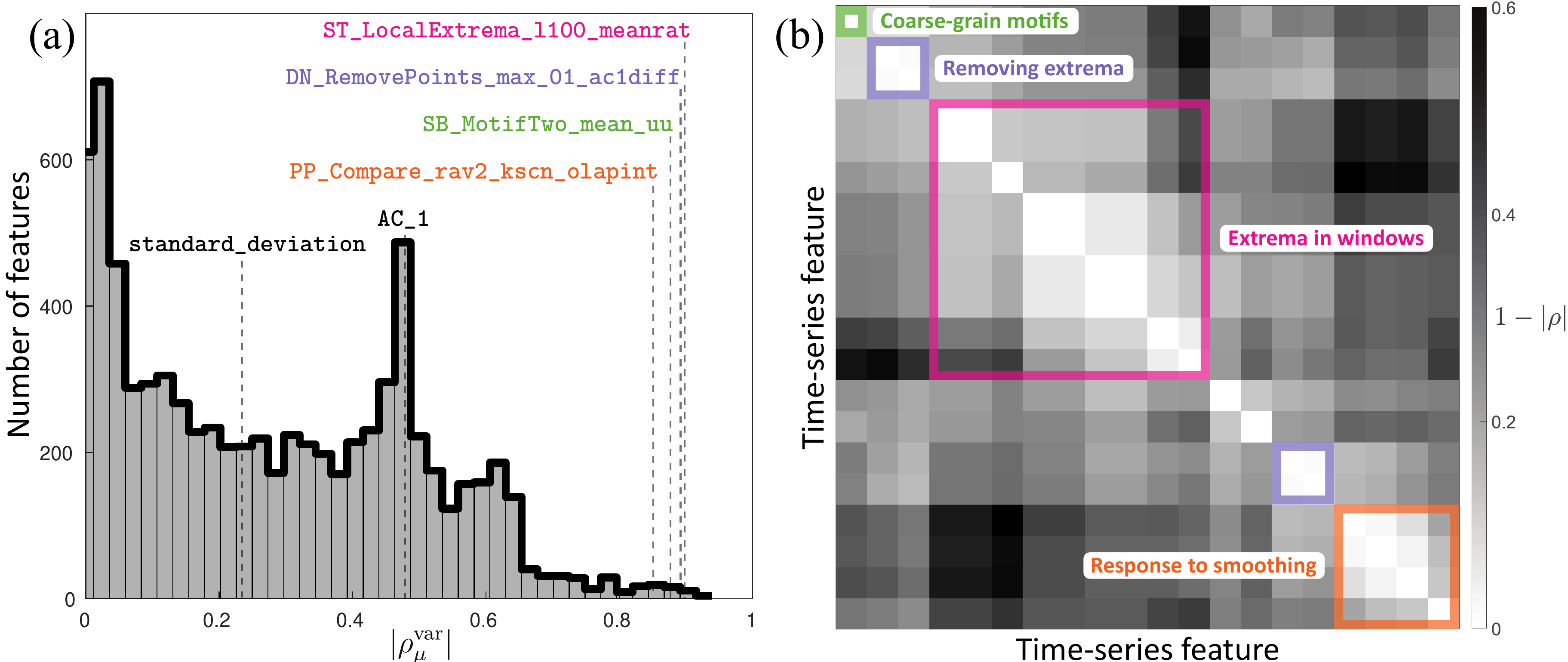}
	{\phantomsubcaption\label{fig:VariableNoiseFeatureHistogram}}
	{\phantomsubcaption\label{fig:featureSimilarityVariable}}
	\cprotect\caption{In the variable-noise setting, we identified a set of high-performing time-series features that measure properties of the distribution of time-embedded values.
	(a) A histogram of the variable-noise scores, \absrhovar{}, is shown across all $7492$ \textit{hctsa} time-series features.
	Conventional indicators of criticality perform poorly, including standard deviation (\rhovar{} $= 0.23$, annotated as \verb|standard_deviation|) and lag-1 autocorrelation (\rhovar{} $= 0.48$, annotated as \verb|AC_1|).
	(b) The top 20 features in the variable-noise case (\absrhovar{} $> 0.86$) are plotted in a pairwise correlation matrix, revealing many diverse clusters of features.
    Families of conceptually similar features are annotated, and the top features from outlined clusters were selected for closer study.
	Features annotated in (a) are detailed in \cref{tab:glossary}, and a sorted list of the features in (b) is in Supplemental Table~S3.}
    \label{fig:VariableNoisePanels}
\end{figure*}

\subsection{The fixed-noise case} \label{sec:FixedNoise}

We first investigated the fixed-noise case, where individual features are scored according to their correlation with $\mu$ for a fixed noise level $\eta$.
We quantified this correlation as \rhofix{} (an average of absolute Spearman correlation coefficients across each of 100 noise levels, $\eta$, cf. \cref{sec:FeatureScoring}).
We hypothesized that features that are well-known to track the vicinity of a system to a critical point---including measures of signal variance and autocorrelation---would receive high \rhofix{} scores.
We also sought to investigate whether any new types of features show strong performance.

The distribution of \rhofix{} values across all 7492 time-series features is shown as a histogram in \cref{fig:FixedNoiseFeatureHistogram}.
We find many time-series features with strong performance on this fixed-noise task, e.g., 1348 of the 7492 tested features have \rhofix{} $ > 0.8$, and some features were scored as high as \rhofix{} $= 0.98$.
To more closely investigate the highest performing time-series features, we focused on the 100 features with \rhofix{} $ > 0.97$.
We then sought to isolate groups of similarly behaving features from within this set of 100 features by computing pairwise Spearman correlation coefficients between all pairs of features (calculated using all time series), and reordering them using hierarchical linkage clustering.
The results, shown in \cref{fig:featureSimilarityFixed}, reveal two distinct groups of high-performing features: one group measuring autocorrelation properties (annotated with an orange square), and another (smaller) group measuring properties of the distribution of time-series values, including mean and variance (annotated with a blue square).

The first cluster contains features that are sensitive to autocorrelation; all features in this cluster are highly correlated to lag-1 autocorrelation (\textit{hctsa} feature name: \verb|AC_1|), which has \rhofix{} $ = 0.97$.
Disregarding the confounds of \noisestrength{} and the time step, lag-1 autocorrelation and other features that measure timescales of a system are sensitive to the critical slowing down that occurs close to the critical point~\cite{Dakos2008}.
As well as lag-1 autocorrelation, our data-driven analysis also reveals a range of conceptually related features that can also effectively capture the same variation in the self-correlation timescale, including measures of automutual information~\cite{Thomas2006}, properties of fitted autoregressive (AR) time-series model residuals, and others (see Supplemental Table S2 for a list of the top $100$ fixed-noise features, clustered by similarity).

In the second cluster of features, which are related to the distribution of time-series values, the standard deviation (\textit{hctsa} feature name: \verb|standard_deviation|) displays very strong performance, with \rhofix{} $ = 0.98$.
This is consistent with expectation---the potential function, described in \cref{sec:Potential}, flattens when approaching a critical point, leading to higher-variance fluctuations for a given noise level $\eta$.
Moreover, since the potential for our model system has a reflecting boundary at $x = 0$ [see \cref{eqn:SupercriticalHopfPotential}], we also find that measures of central tendency such as the mean, median, and root-mean-square, are highly correlated to the spread of the distribution (as captured by the standard deviation).

Our results have thus flagged two types of features that vary with the DTC, recapitulating prior literature which has focused on critical slowing down (captured by measures of time-series autocorrelation~\cite{Dakos2008}, including the simple and effective indicator, lag-1 autocorrelation~\cite{Scheffer2009a, Scheffer2012, Kuehn2011, Kuehn2013, Dakos2012}) and fluctuations of increased variance (captured by the spread of the distribution of time-series values) \cite{Carpenter2006} near a critical point.
A complete list of \textit{hctsa} features along with their fixed-noise scores, \rhofix{}, is in Supplemental Table S1.

\subsection{The variable-noise case}
\label{sec:variablenoise}

Having established the ability of our data-driven approach to recapitulate a theoretical literature on time-series features for tracking the DTC in a system with fixed noise, we next investigated the more difficult problem of finding features that track the DTC in a variable-noise setting.
We are unaware of prior work that has examined this relevant real-world scenario, which involves identifying statistical properties of time series that are sensitive to $\mu$ but insensitive to $\eta$.
The lack of an \textit{a priori} understanding of how to construct a noise-level-robust indicator of the DTC from finite time series, and the difficulty (or intractability) of a direct analytical route to tackling the problem, makes it an ideal setting for our data-driven approach.

\begin{figure*}[hp!]
\vspace{-1em}
\centering
{\phantomsubcaption\label{fig:featurescatters_standard_deviation}}
{\phantomsubcaption\label{fig:featurescatters_AC}}
{\phantomsubcaption\label{fig:featurescatters_DN_RemovePoints}}
{\phantomsubcaption\label{fig:featurescatters_SB_MotifTwo}}
{\phantomsubcaption\label{fig:featurescatters_PP_Compare}}
{\phantomsubcaption\label{fig:featurescatters_ST_LocalExtrema}}
{\phantomsubcaption \label{fig:featurescatters_fitFeature}}
{\phantomsubcaption \label{fig:featurescatters_NewFeature}}
    \includegraphics[width=\textwidth]{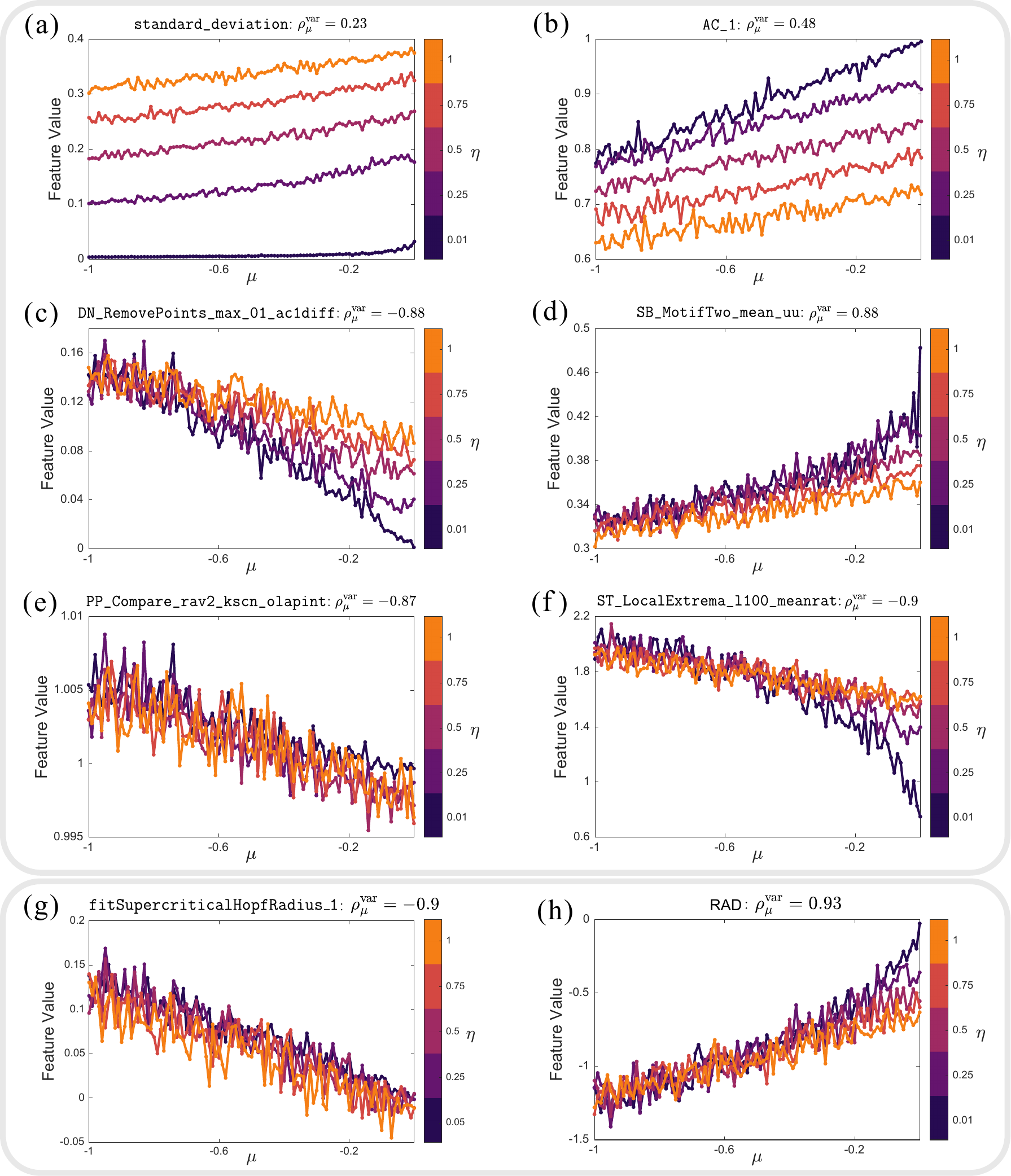}
	{\cprotect \caption{Conventional features for tracking the DTC only perform well in the fixed-noise setting, but we uncover new features that robustly track the DTC across different \noisestrength{}s.
    Feature values are plotted against $\mu$ for five values of the \noisestrength{} $\eta = 0.01, 0.25, 0.5, 0.75, 1$.
	Two of the top fixed-noise features, (a) standard deviation and (b) lag-1 autocorrelation, are highly correlated to $\mu$ for any given value of $\eta$ (the fixed-noise case), but are poorly correlated across multiple $\eta$ values (the variable-noise case).
	(c)--(f) The highest-performing time-series features in the variable-noise case are highly correlated to $\mu$ both within and across $\eta$ values, demonstrating robust tracking of the DTC despite unknown \noisestrength{}.
    (g) A new feature introduced here, \texttt{fitSupercriticalHopfRadius\_1}, distils the key algorithmic elements of the top-performing features.
    However, the curve-fitting algorithm is unstable for $\eta < 0.05$, so only $\eta = 0.05$ is shown for this feature.
    (h) A second new feature, \newfeature{}, uses elementary time-series operations to yield a computationally efficient and numerically stable estimate of the DTC.
    Each feature is summarized in \cref{tab:glossary} and detailed in \cref{sec:UnderstandingTopFeatures}.
	}
 \label{fig:featurescatters}}
\end{figure*}

We computed the variable-noise performance score, \rhovar{}, for all candidate time-series features in \textit{hctsa} (full results are in Supplemental Table S1).
The distribution of \absrhovar{} across all 7492 features is plotted in \cref{fig:VariableNoiseFeatureHistogram}, including annotated scores of two selected features (the standard deviation and lag-1 autocorrelation).
Reflecting the increased difficulty of the variable-noise setting, \absrhovar{} values are lower on average than \rhofix{}.
We nevertheless observed a tail of high-scoring features; e.g., 49 time-series features have \absrhovar{} $> 0.8$.
As expected, the top-performing features in the fixed-noise case, including lag-1 autocorrelation (\verb|AC_1|) and standard deviation (\verb|standard_deviation|) are sensitive to both $\mu$ and $\eta$, and thus do not accurately track variation in $\mu$ in the presence of confounding variation in $\eta$.
For example, for \verb|standard_deviation|, \rhofix{} $= 0.98$ (in the fixed-noise setting) drops to \rhovar{} $= 0.23$ (in the variable-noise setting), while for \verb|AC_1|, \rhofix{} $= 0.97$ drops to \rhovar{} $= 0.48$.
To understand this behavior, we plotted the dependence of these two features as a function of $\mu$ for five selected noise levels in \cref{fig:featurescatters_AC,fig:featurescatters_standard_deviation}.
We see that these features vary monotonically with $\mu$ for any given $\eta$ value (underlying their strong fixed-noise performance), but they are highly sensitive to variation in $\eta$, making them unreliable indicators of the DTC in the variable-noise setting.

We next aimed to better understand the high-performing features that make up the right tail of \cref{fig:VariableNoiseFeatureHistogram}.
Focusing on the top twenty features, with \rhovar{}$> 0.86$, we plotted their pairwise distance matrix (on Spearman correlation distances, $1 - |\rho|$) in \cref{fig:featureSimilarityVariable} (see Supplemental Table S3 for details on each cluster).
One large family of highly correlated features examines how properties of extreme values within small windows (of ${\sim}100$ samples) are distributed across the time series, such as the average ratio between the maximum and minimum value in each window [\verb|ST_LocalExtrema_l100_meanrat|, shown in pink in \cref{fig:FixedNoiseFeatureHistogram}].
Other high-performing features include those that measure the change in autocorrelation after a proportion of time series extrema are removed (such as \verb|DN_RemovePoints_max_01_ac1diff|, purple), the occurrence of simple motifs in a symbolization of the time series (i.e., transforming a sequence of real values to a symbolic string; \verb|SB_MotifTwo_mean_uu|, green), or the change in the distribution of time-series values after smoothing the time series using a moving average (\verb|PP_Compare_rav2_kscn_olapint|, orange).

\subsection{Algorithmic steps underlying noise-robust features}
\label{sec:UnderstandingTopFeatures}
\label{sec:Potential}
\label{sec:invariantdensity}
\label{sec:Sigma}

In the previous section, we isolated a set of time-series features that can track the control parameter $\mu$, while being minimally affected by changes in the noise level $\eta$, with a surprisingly strong correlation (\rhovar{} $> 0.86$).
In this variable-noise setting, conventional indicators of the DTC, like standard deviation and lag-1 autocorrelation, perform poorly.
In this section we aim to understand why these features perform so well.
We are able to explain this by inspecting the algorithms underlying four of the highest-performing features, selected to represent the main clusters of high-performing feature behavior [in \cref{fig:featureSimilarityVariable}].
These four features, which are named and briefly described in \cref{tab:glossary}, are annotated on \cref{fig:FixedNoiseFeatureHistogram,fig:VariableNoiseFeatureHistogram}.
Scatter plots of these top features with the DTC $\mu$ for different noise levels $\eta$ (plotted in~\cref{fig:featurescatters_DN_RemovePoints,fig:featurescatters_PP_Compare,fig:featurescatters_SB_MotifTwo,fig:featurescatters_ST_LocalExtrema}), as well as the joint distribution of \textit{hctsa} features over the fixed-noise and variable-noise scores (plotted in \cref{fig:fixedvariableScatter}), indicate their high performance in both the fixed-noise and variable-noise cases.
Unlike the standard deviation [\cref{fig:featurescatters_standard_deviation}] and lag-1 autocorrelation [\cref{fig:featurescatters_AC}], these high-performing features vary strongly with $\mu$ in a similar way across noise levels, $\eta$, demonstrating their robustness to noise.

Despite appearing to be distinct features, close inspection of the algorithmic steps underlying each top feature revealed two key shared algorithmic components:
(i) they involve estimating a statistic derived from the distribution of time-series values, such as a proportion of points within an interval of values; and
(ii) they compare this statistic to the standard deviation of the incrementally differenced time series (a quantity that is closely related to the lag-1 autocorrelation).
In this section we aim to summarize how these two algorithmic steps, which appear critical to the top-performing features, allow them to act as noise-robust estimators of the DTC.
As with conventional metrics of criticality, such as standard deviation and autocorrelation, we notice that both algorithmic steps are closely related to a system's potential function.
By expressing these two properties in terms of the potential, we are able to explain how these features can successfully infer the DTC despite uncertain \noisestrength{}.
We first derive the relationship between the potential and the two algorithmic steps: (i) the distribution of values; and (ii) the spread of differences.

\paragraph{The shape of the potential $V(x)$ depends on $\mu$.}
The ability of some features to accurately track the DTC in the presence of uncertain \noisestrength{} can be understood in terms of the potential function of a system.
The potential function of a stochastic system describes the effect of the deterministic components on the system dynamics: the gradient of the potential at a given point determines the rate of change due to the deterministic terms at that point~\cite{Strogatz1994}.
In this formalism, the evolution of a system can be viewed intuitively as the trajectory of a heavily damped particle following the gradient of the potential toward a local minimum~\cite{Strogatz1994}, in addition to any diffusive (or stochastic) drives.
The potential formulation provides a useful way of understanding critical slowing down and other conventional metrics of the DTC~\cite{Scheffer2012}.
The potential function for the model system studied here is obtained by integrating \cref{eqn:SupercriticalHopfRadial}, giving
\begin{equation}
\label{eqn:SupercriticalHopfPotential}
V(x;\mu) = -\frac{\mu x^2}{2} + \frac{x^4}{4}\,,\quad x \ge 0\,.
\end{equation}

\begin{figure*}[ht!]
    \centering
    \includegraphics[width=\textwidth]{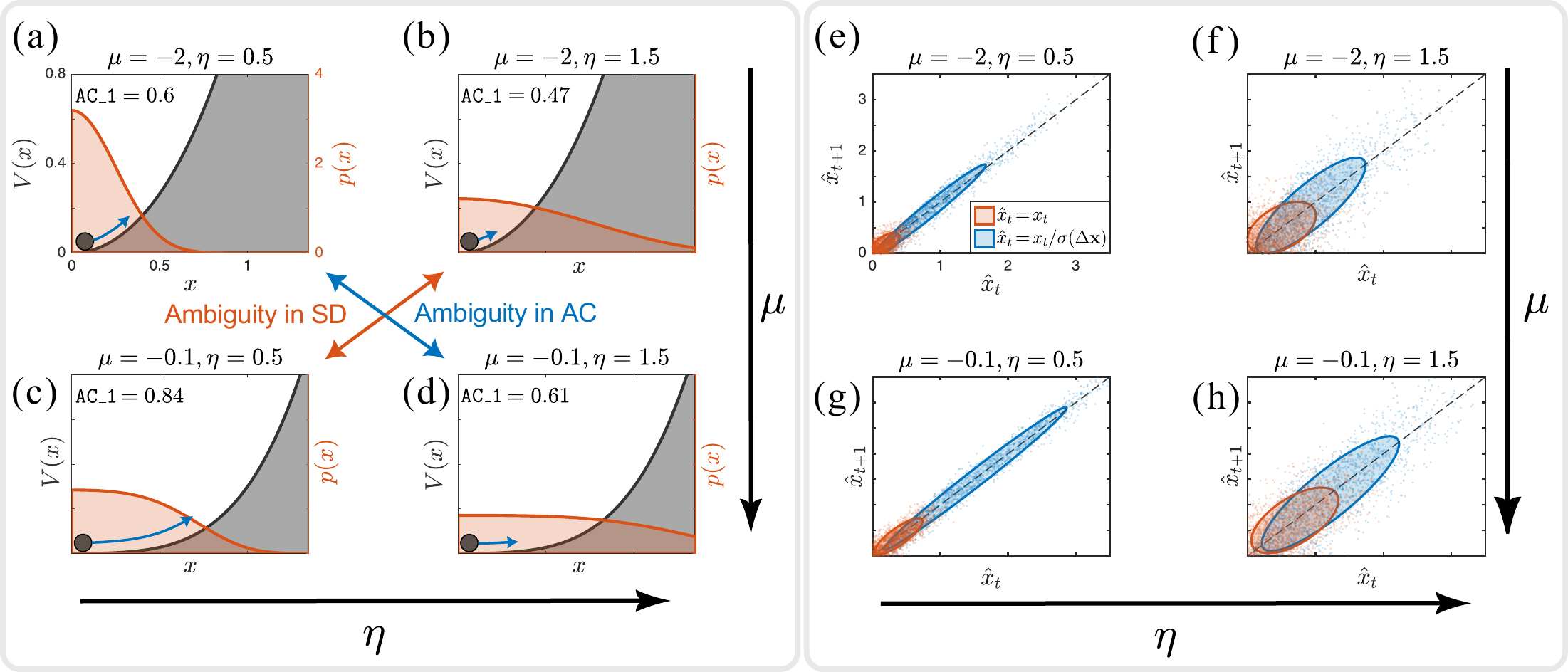}
    {\phantomsubcaption\label{fig:TheorySchematic_a}}
    {\phantomsubcaption\label{fig:TheorySchematic_b}}
    {\phantomsubcaption\label{fig:TheorySchematic_c}}
    {\phantomsubcaption\label{fig:TheorySchematic_d}}

    {\phantomsubcaption\label{fig:TheorySchematic_e}}
    {\phantomsubcaption\label{fig:TheorySchematic_f}}
    {\phantomsubcaption\label{fig:TheorySchematic_g}}
    {\phantomsubcaption\label{fig:TheorySchematic_h}}
    \cprotect\caption{Rescaling time-series values by the spread of differenced values corrects for the confounding effect of a variable \noisestrength{}.
    (a)--(d) We plot the potential function $V(x)$ (determined by the control parameter, $\mu$, shown black), and distribution of values $p(x)$ (orange) for four combinations of $\mu$ and $\eta$: (a) $\mu = -2$, $\eta = 0.5$, (b) $\mu = -2$, $\eta = 1.5$, (c) $\mu = -0.1$, $\eta = 0.5$, and (d) $\mu = -0.1$, $\eta = 1.5$.
    The control parameter $\mu$ defines the DTC and determines the potential function, $V(x)$.
    The flatness of $V(x)$ gives rise to critical slowing down that, in combination with the \noisestrength{} $\eta$, determines $p(x)$ and the lag-1 autocorrelation (annotated as \AC{}, top right).
    Both $p(x)$ and lag-1 autocorrelation values give ambiguous estimates of the DTC: different pairs of $\mu$ and $\eta$ can result in the same feature value.
    (e)--(h) Scatters of $(x_t, x_{t+1})$ (orange) and the rescaled $(x_t/\sigma(\Delta \mbf{x}), x_{t+1}/\sigma(\Delta \mbf{x}))$ (blue) for the same four parameter settings as in (a)--(d):
    (e) $\mu = -2$, $\eta = 0.5$, (f) $\mu = -2$, $\eta = 1.5$, (g) $\mu = -0.1$, $\eta = 0.5$, and (h) $\mu = -0.1$, $\eta = 1.5$.
    After rescaling by the standard deviation in the $x_{t+1} - x_t$ direction, $\sigma (\Delta \mbf{x})$ (described in \cref{sec:NewFeature}), the distribution of $x_t + x_{t+1}$ becomes less sensitive to changes in the \noisestrength{}.
    Annotated ellipses indicate the covariance of $(x_t, x_{t+1})$ (orange), which is strongly affected by changes in $\eta$, but becomes more characteristic of $\mu$ after rescaling by $\sigma (\Delta \mbf{x}) $ (blue).
    }
	\label{fig:TheorySchematic}
\end{figure*}

As depicted in \crefrange{fig:TheorySchematic_a}{fig:TheorySchematic_d}, as $\mu$ increases toward $0$, $V(x)$ becomes shallower, reducing the restorative force toward the origin.
This change in the shape of the potential results in dynamics with slower fluctuations (increased autocorrelation) and diffusion over a larger domain [increased standard deviation, cf. orange distributions in \crefrange{fig:TheorySchematic_a}{fig:TheorySchematic_d}].
However, \crefrange{fig:TheorySchematic_a}{fig:TheorySchematic_d} also show that $\eta$ can make the relationship between the potential and autocorrelation or standard deviation ambiguous: the time-series autocorrelation takes a similar value for $\mu = -2, \eta = 0.5$ [\cref{fig:TheorySchematic_a}] as for $\mu = -0.1, \eta = 1.5$ [\cref{fig:TheorySchematic_d}], and the distribution has a similar standard deviation for $\mu = -2, \eta = 1.5$ [\cref{fig:TheorySchematic_b}] and $\mu = -0.1, \eta = 0.5$ [\cref{fig:TheorySchematic_c}].
While both the potential function and driving noise determine how a system evolves over time, we can see that a high-performing algorithm for estimating $\mu$ in the presence of variable noise should target a property that is specifically affected by $\mu$: the shape of the potential function, $V(x;\mu)$.

\paragraph{The invariant density $p(x)$.}
Under the conceptual framing provided by the potential formulation, we next aimed to investigate whether the top-performing features are able to extract a noise-robust estimate of $\mu$ from time-series data by estimating the shape of the potential, $V(x;\mu)$.
As described above, all top-performing features measure properties of the distribution, suggesting that this step may be relevant to robustly estimating the DTC (see \cref{sec:variablenoise}).
Following this connection, we investigated the invariant density $p(x)$: the probability density to which the distribution of values from a stationary system will converge over time.
The invariant density is valuable for describing bifurcations when the abruptness of the qualitative change that occurs in the deterministic system is destroyed or smoothed by noise~\cite{Meunier1988}.
For a stationary system, the invariant density can be derived from the potential function and the noise term by taking the Fokker--Planck equation in the limit of infinite time~\cite{Schumaker1987}.
For the system studied here, \cref{eqn:SupercriticalHopfRadial}, the invariant density has the form
\begin{equation}
\label{eqn:InvariantDensity}
p(x; \mu, \eta) = A\exp(\frac{-2V(x; \mu)}{\eta^2})\,, \quad x \ge 0\,,
\end{equation}
which is normalized to unit probability mass by $A$~\cite{Meunier1988}.
For fixed $\eta$, the potential can be inferred from the invariant density as
\begin{equation}
    \label{eqn:PotentialDensity}
    V(x;\mu) = -\frac{1}{2}\eta^2\left[\ln p(x) - \ln A\right]\,,
\end{equation}
where it is straightforward to obtain an estimate of the probability density, $p(x)$, from measured data (e.g., using a kernel estimator~\cite{Chen2017}).
Since \cref{eqn:SupercriticalHopfPotential}, and therefore \cref{eqn:PotentialDensity}, depend on $\mu$, time-series features that measure properties of the distribution of values in the time series (such as mean, standard deviation, and skewness) are able to provide accurate estimates of the DTC in the fixed-$\eta$ case (as we verified in Sec.~\ref{sec:FixedNoise}).
However, as \cref{eqn:PotentialDensity} depends strongly on $\eta$, features measuring the distribution of time-series values are highly sensitive to noise and thus have low \rhovar{} (as we found in Sec.~\ref{sec:variablenoise}).

We now seek to incorporate the second algorithmic component of the top-performing features, the spread of differences, into our expression for the potential function, in the hope that it may eliminate the \noisestrength{} $\eta$ from the invariant density and yield a noise-insensitive estimate of $\mu$.

\paragraph{The spread of differences, $\sigma(\Delta \mbf{x})$.}
In addition to the distribution of time-series values, the time-series features with the highest \rhovar{} also indirectly measure the spread of the incrementally differenced time series $\Delta \mbf{x}$, where $\Delta$ is the first difference operator and $\Delta x_t = x_t - x_{t-1}$~\cite{Hamilton1994}.
We write this spread here as $\sigma(\Delta \mbf{x})$, where $\sigma$ is the standard deviation (see \cref{sec:TopFeaturesSupplement} for further details).
Considering the drift and diffusion terms of \cref{eqn:SupercriticalHopfRadial}, we see that the standard deviation of increments of the Wiener process scale with $\sqrt{\Delta t}$~\cite{Bass2011}, whereas increments of $(\mu x - x^3)dt$ scale with $\Delta t$.
Hence for a small time step $\Delta t$ the standard deviation of the differenced time series is dominated by the stochastic component $\eta dW$.
In the short-timestep limit, $\Delta t \to 0$, such that $\sqrt{\Delta t} \gg \Delta t$, we obtain an approximation for inferring $\eta$:
\begin{equation} \label{eqn:sigma}
    \sigma(\Delta \mbf{x}) \approx \eta \sqrt{\Delta t}\,,
\end{equation}
where $\sigma(\Delta \mbf{x})$ is the standard deviation of the incrementally differenced time series $\Delta\mbf{x}$.

The top-performing variable-noise features all involve measuring this dynamical quantity, $\sigma(\Delta \mbf{x})$, that is informative of $\eta$.
This insight allows us to understand how they may be able to eliminate the \noisestrength{} term from the invariant density to more accurately estimate the DTC.
Incorporating this empirical estimate of $\eta$, from Eq.~\eqref{eqn:sigma}, into our expression for the potential in terms of the invariant density, Eq.~\eqref{eqn:PotentialDensity}, yields
\begin{equation}
\label{eqn:PotentialDifferenceDensity}
    V(x;\mu) \approx -\frac{\sigma^2(\Delta \mbf{x})}{2\Delta t}\left[\ln p(x) - \ln A \right]\,,
\end{equation}
where we have assumed a fixed sampling period $\Delta t$.

To test \cref{eqn:PotentialDifferenceDensity}, we used it to infer the DTC in the fixed-noise and variable-noise cases studied here.
For the system given by \cref{eqn:SupercriticalHopfRadial}, we can infer $V(x) \Delta t$, and therefore the DTC, using two properties that are easily measured from time-series data: the probability density $p(x)$ and the spread of differences $\sigma (\Delta \mbf{x})$.
The algorithmic procedure for inferring the DTC from data using \cref{eqn:PotentialDifferenceDensity} is as follows:
(i) produce a kernel-density estimate for $p(x)$;
(ii) calculate $\sigma(\Delta \bf{x})$ using the standard deviation and the first difference operator~\cite{Hamilton1994};
(iii) transform the density to a potential using \cref{eqn:PotentialDifferenceDensity};
(iv) curve-fit \cref{eqn:SupercriticalHopfPotential} to the estimated potential;
then (v) infer $\mu$ by calculating the value of the second derivative at $x = 0$.
We give this procedure the feature name \verb|fitSupercriticalHopfRadius_1|, and show how it varies with both $\mu$ and $\eta$ in \cref{fig:featurescatters_fitFeature}.

Since \verb|fitSupercriticalHopfRadius_1| varies strongly with the control parameter $\mu$ but minimally with the \noisestrength{} $\eta$, we find that it tracks the DTC in the variable-noise case far more strongly than conventional fixed-noise metrics, yielding scores that are on-par with the top-performing \textit{hctsa} features (see \cref{sec:variablenoise}): \rhofix{} $= 0.92$ and \rhovar{} $= 0.90$ (see Supplemental Table~S1 for a complete list of scores for both new features and \textit{hctsa} features).
The strong performance of \verb|fitSupercriticalHopfRadius_1| suggests that \cref{eqn:PotentialDifferenceDensity} is indeed capturing key algorithmic principles relevant to robustly tracking the DTC.
However, the top \textit{hctsa} features (detailed in \cref{sec:TopFeaturesSupplement}) are able to infer the DTC with disparate and often simple algorithms that do not rely on the computationally expensive steps of density estimation and curve fitting, and do not require explicitly knowing the form of the potential or the noise process.
Furthermore, we found that \verb|fitSupercriticalHopfRadius_1| suffers from numerical instability and produces noisy values when the \noisestrength{} is low, since the distribution is tightly concentrated around $x \approx 0$; in our simulations, this instability only occurred at $\eta = 0.01$.
Armed with a theoretical explanation of how time-series features are able to estimate the DTC in the presence of noise, we next aimed to develop a simple, efficient, stable, and generic algorithm that implements the potential-based inference of the DTC described above.

\subsection{Rescaled Auto-Density: a new noise-robust metric of the distance to criticality}
\label{sec:NewFeature}

In this section, we develop and test a new feature that directly implements the key algorithmic steps underpinning the success of the top-performing \textit{hctsa} features, allowing us to efficiently estimate the DTC from data.
We first detail a simplified approach to inferring the potential function $V(x)$ [cf. \cref{eqn:PotentialDifferenceDensity}] that eliminates $\eta$ by rescaling time-series values by the standard deviation of incremental time-series differences, $\sigma(\Delta \mbf{x})$.
We then describe how the density of points in an $(x_t, x_{t+1})$ time-delay embedding can be used to infer the DTC from the noise-robust rescaled potential.
Finally, we introduce the Rescaled Auto-Density (RAD), a new time-series feature that uses elementary time-series operations to implement these key algorithmic principles.

\paragraph{Rescaled potential.}
We first notice that when the system is close to the origin, $x \ll 1$, the quadratic term in \cref{eqn:SupercriticalHopfPotential}, $-\mu x^2/2$, dominates the quartic term $x^4/4$.
Since our system is only driven by noise and has no extreme jumps or perturbations, it is naturally confined to a region close to the origin by the `steep walls' of the quartic term [see \crefrange{fig:TheorySchematic_a}{fig:TheorySchematic_d}].
This behavior can be observed from the Taylor expansion of the invariant density, \cref{eqn:InvariantDensity}, given by
\begin{equation}
\label{eqn:TaylorMade}
\tilde{p}(x; \mu, \eta) = B\left( 1 + \mu\frac{x^2}{\eta^2} + \frac{\mu^2}{2}\frac{x^4}{\eta^4} - \frac{x^4}{2\eta^2} + \mathcal{O}(x^6)\right)\,,
\end{equation}
where $B$ is a constant that normalizes $\tilde{p}(x)$ to unit probability mass.
Note, however, that this behavior breaks down when $\eta$ becomes too large (the noise drives the system to extreme values where the quartic term dominates) or $\mu$ becomes too small (the quartic term dominates as the quadratic term is silenced).
Under the approximation that the system tends to reside close to equilibrium, $x \ll 1$, we can write a simplified, rescaled potential as
\begin{equation}
\label{eqn:PotentialApproximation}
\hat{V}(\hat{x};\mu) = -\frac{\mu \hat{x}^2}{2} = -\frac{\ln[p(\hat{x})] - c}{2\Delta t} \,,
\end{equation}
where $\hat{x} \ge 0$, $c$ is a constant, and $\hat{x} = x/[\sigma(\Delta \mbf{x})] \approx x/(\eta \Delta t)$.
This approximate potential, $\hat{V}$, provides an algorithmically simpler way to estimate the DTC: we first rescale the system with $\hat{x} =x/[\sigma(\Delta \mbf{x})]$ (aiming to eliminate the contribution of $\eta$) and then infer the potential of the rescaled system using the distribution of the rescaled values (to estimate $\mu$).
Unlike the theory described above, in Sec.~\ref{sec:UnderstandingTopFeatures}, this simplified, highly approximate approach is not specific to the potential function of our model system.

\paragraph{The auto-density.}
Having identified a simpler method for `sensing' the underlying potential function governing the deterministic dynamics, by rescaling the system with the standard deviation of incremental time-series differences, we now develop a simple algorithm for estimating the DTC from the inferred potential.
We notice that the key attributes for robustly inferring the DTC---the invariant density and the spread of differences---can both be measured from the distribution of time-series values in an $(x_{t}, x_{t+1})$ time-delay embedding~\cite{takens2006detecting}, as illustrated in \crefrange{fig:TheorySchematic_e}{fig:TheorySchematic_h}.
For the purposes of naming our feature, we refer to this two-dimensional distribution in $(x_{t}, x_{t+1})$ as the lag-1 \textit{auto-density} (in analogy to autocorrelation).
This auto-density captures the two properties we have found to be key for robustly inferring the DTC: the spread of differences, $\sigma(\Delta \mbf{x})$, and the distribution, $p(x)$.
First, in a two-dimensional embedding space, $(x_{t},x_{t+1})$, the factor $\sigma(\Delta \mbf{x})$ is proportional to the standard deviation of shortest distances from $x_t = x_{t+1}$, or $x_{t+1} - x_t$.
Second, when the sampling period is small, $\Delta t \ll 1$, the distribution $p(x)$ is well approximated by the linear projection of $x_t$ values onto the line defined by $x_t = x_{t+1}$, proportional to the distribution of the quantity $x_t + x_{t+1}$.
The approximation given by \cref{eqn:PotentialApproximation} can then be implemented by rescaling $x_t$ and $x_{t+1}$ with the width of the auto-density in the $x_{t+1} - x_t$ direction.
As illustrated in \cref{fig:TheorySchematic_e,fig:TheorySchematic_f,fig:TheorySchematic_g,fig:TheorySchematic_h}, rescaling the time-series values produces a distribution that is minimally sensitive to changes in the \noisestrength{}.

\paragraph{The rescaled auto-density.}
We now encapsulate the key algorithmic steps described above into a new time-series feature, \newfeature{}, that aims to robustly infer the DTC of noisy systems.
It begins by partitioning the time-series values about a threshold that is insensitive to changes in extreme densities.
The median ($\Tilde{{x}})$ is a suitable choice for this threshold, partitioning the time series into two sets: time-series values above the median
\begin{equation}
U = \{x_t: x_t \geq \tilde{x}\}\,,
\end{equation}
and the time-series values below the median
\begin{equation}
L = \{x_t: x_t < \tilde{x}\}\,.
\end{equation}
\newfeature{} then aims to summarize of the shape of the invariant density while avoiding the need for the curve fitting (as required by \verb|fitSupercriticalHopfRadius_1|, cf. \cref{sec:Sigma}).
For a simple measurement of the shape of the probability density, we quantify the tailedness as the difference in the average density between the upper and lower partitions.
Here we define the average density in each partition as the ratio of the probability mass contained in the partition and the width of the partition, measured by the standard deviation.
Noting that the two partitions are split by the median, and therefore have an equal probability mass of $0.5$, we can capture the average density in a partition as the inverse of the standard deviation, $1/\sigma$.
Rescaling the difference between the average densities of the upper and lower partitions (which measures the shape of the distribution, in particular the higher-order moments) by $\sigma(\Delta \mbf{x})$ completes our \newfeature{} feature, which is then given by
\begin{equation}
\label{eqn:NewFeature}
f_\textrm{\newfeature{}} = \sigma(\Delta \mbf{x}) \left[\frac{1}{\sigma(U)} - \frac{1}{\sigma(L)} \right]\,.
\end{equation}
\newfeature{} takes negative values far from criticality and approaches zero when a system is close to the critical point. %
We also define the centered \newfeature{}, denoted as c\newfeature{}, which is applicable to non-radial data:
\begin{equation}
    \label{eqn:centerednewfeature}
    f_\textrm{c\newfeature{}} = f_\textrm{\newfeature{}}(\abs{x - \tilde{x}})\,,
\end{equation}
where $\tilde{x}$ is the median value of $\mbf{x}$ and $\abs{\cdot}$ is the absolute value. %
To verify the DTC-tracking performance of \newfeature{}, we applied it to the fixed-noise and variable-noise cases analyzed above, as shown in \cref{fig:featurescatters_NewFeature}.
In the variable-noise case, \newfeature{} out-performed all other \textit{hctsa} features, with \rhofix{} $= 0.93$ (see the RAD feature in Supplemental Table S1).
\newfeature{} also exhibits competitive performance relative to the top features in the fixed-noise setting, with \rhovar{}$= 0.93$.
Furthermore, we verified that \newfeature{} out-performs conventional metrics in a range of other systems with various normal forms and noise processes (see \cref{sec:robustnesssupplement}).
\newfeature{} is a straightforward and transparent algorithm that behaves as a reliable indicator of DTC in noisy systems.
Implementations of \newfeature{} and centered \newfeature{} are provided in an accompanying Matlab code repository~\cite{Harris2023}, and are also included in the \textit{hctsa} time-series feature library~\cite{Fulcher2017}.


\begin{figure*}[htbp!]
    \includegraphics[width=\textwidth]{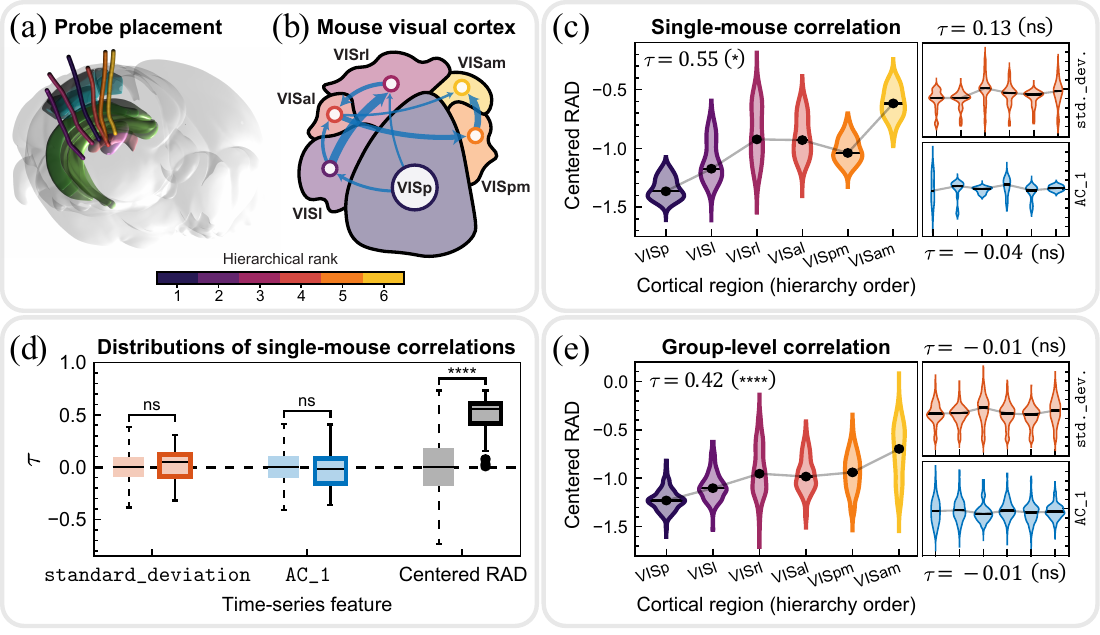}
    {\phantomsubcaption\label{fig:neuropixelsviz}}
	{\phantomsubcaption\label{fig:hierarchyviz}}
	{\phantomsubcaption\label{fig:corticalcorrelations_individual}}
 	{\phantomsubcaption\label{fig:corticalcorrelations_distribution}}
	{\phantomsubcaption\label{fig:corticalcorrelations_pooled}}
	\cprotect\caption{\newfeature{} tracks the anatomical hierarchy of mouse visual cortical regions from neural electrophysiology data.
    (a) Six Neuropixels probes (lines, black to yellow) in the mouse brain and the key areas they intersect (visual cortex, blue; hippocampus, green; and thalamus, pink).
	(b) Six visual cortical regions and their hierarchy ranks (black to yellow) shown with anatomical similarities from from~\citet{Siegle2021} (blue arrows, width weighted by similarity, directed from lower-order to higher-order areas). %
	(c) The distribution of centered \newfeature{}, over $n \in [15, 24] $ channels, for each of the six cortical regions in a single representative mouse (session $1065908084$). The strong Kendall's $\tau$ correlation of centered \newfeature{} to the hierarchy score ($\tau = 0.55$, $p = 0.03$; *) is annotated, and is equal to the group median---see (d).
    Connected dots show the median centered \newfeature{}, and regions are ordered based on the anatomical hierarchy shown in (b).
    Outside right are similar plots for standard deviation (red, top right; $\tau=0.13$, $p=0.4$) and lag-1 autocorrelation, \AC{} (blue, bottom right; $\tau = -0.04$, $p=0.7$), both showing weak and non-significant (ns) correlations.
	(d) The distribution over mice ($n=39$), of Kendall's $\tau$ correlations, as depicted by boxplots in (b), for standard deviation (red), lag-1 autocorrelation (blue), and centered \newfeature{} (gray). Bars show median values, boxes show the interquartile range, dots show outliers ($>1.5\times$IQR from the nearest quartile), and whiskers show the range of non-outlier values. %
    $p$-values for Mann-Whitney U tests between the correlations for each feature and their corresponding null distributions (shown by dashed boxplots, over $n=39 \times 10^{6}$ values of $\tau$ (see main text) are annotated.
    Single-mouse correlations of centered RAD to the hierarchical level are highly significant, with $p < 10^{-18}$ (****), but are not significant for both standard deviation ($p=0.3$) and lag-1 autocorrelation ($p = 0.5$).
    (e) The distribution of centered \newfeature{} pooled across all mice and channels (yielding $n\approx 750$ data points for each region). The group-level Kendall's $\tau$, computed on this pooled data, is annotated alongside an estimated $p$-value (see main text).
    Even at a group level, the correlation of centered \newfeature{} to the hierarchy rank is strong and highly significant ($\tau = 0.42$, $p<10^{-6}$), unlike both standard deviation (red, top right; $\tau = -0.01$, $p=0.8$) and lag-1 autocorrelation (blue, bottom right; $\tau = -0.01$, $p=0.5$).
	}
	\label{fig:casestudy}
\end{figure*}

\subsection{Structure--function organization across the mouse visual cortex} 
\label{sec:casestudy} %

We next aimed to test the performance of \newfeature{} in a real-world setting using electrophysiological data, alongside conventional metrics of tracking the DTC. %
Real-world complex systems, such as the brain, are thought to exploit near-critical states for computational advantage~\cite{kinouchi2006,Gautam2015,Cocchi2017,Fontenele2019}, with near-critical systems exhibiting enhanced dynamic range~\cite{kinouchi2006,Shew2009,gollo2017}, input separation and sensitivity, information-storage capacity, and information transfer capabilities~\cite{Shew2012,Munoz:2018dn,OByrne2022}.
In neural systems, which benefit from such advantages, evidence for criticality has been found at multiple scales, from single neurons~\cite{gollo2013, gal2013,Meisel2015} to neuronal spike trains~\cite{Wilting2018,Ma2019b,Dahmen2019,Munn2020,Lotfi2021,Morales2023,Jones2023}, local-field potentials~\cite{Shew2015,Hahn2017,Mariani2022}, as well as MEG~\cite{Fusca2023} and BOLD fMRI signals~\cite{Tagliazucchi2012}. %

Numerous studies have suggested that, despite the commonly reported benefits of criticality, the brain tends to operate in a subcritical state~\cite{o2022,priesemann2014,wilting2019}, often associated with focused attention~\cite{tomen2014,fagerholm2015}.
However, other research suggests that the brain fluctuates around a critical threshold~\cite{Fontenele2019}.
A more comprehensive perspective, considering the diversity of brain regions, suggests the advantages of integrating regions with varying dynamical states, combining those closer to criticality with others that are more subcritical~\cite{gollo2017,Hilgetag2020,liang2024}.
Many theories of neural criticality then predict that regions further along the cortical hierarchy (i.e., regions responsible for higher-order cognitive functions such as the integration of complex stimulus features, conscious perception, or decision-making~\cite{DSouza2022}) require longer timescales for integrating information~\cite{Kiebel2008,Murray2014,Chaudhuri2015,Gollo2019}, and are thus expected to sit closer to criticality~\cite{Cocchi2017,Hilgetag2020,liang2024}. %
This hypothesis has been supported, in part and indirectly, by a corresponding hierarchical variation in the excitation-to-inhibition ratio~\cite{DSouza2016,liang2024} and other structural features~\cite{Gollo2015,Fulcher2019}, structural and functional connectivity~\cite{cocchi2016,Glickfeld2017}, as well as intrinsic timescales~\cite{Murray2014,Chaudhuri2015} across cortical areas. %
While much work has analyzed the theoretical and anatomical basis for tuning the DTC of neural systems~\cite{Cocchi2017,Tian2022}, the presence of variable stochastic drives across brain regions---perhaps arising from differences in connectivity to external regions~\cite{Young2021}, nearby vasculature~\cite{Wu2022}, or heterogeneous cytoarchitectures and their influence on high-frequency dynamics~\cite{Shafiei2023}---confound the inference of the DTC in real-world macroscale recordings of neural activity using conventional measures like autocorrelation and variance.

Based on the hypothesis that the proximity of a cortical area to criticality increases along the visual hierarchy~\cite{Cocchi2017,Hilgetag2020,liang2024}, and that brain regions may be subject to different levels of dynamical noise, here we aimed to compare how well \newfeature{} tracks the hierarchical level---measured independently using anatomical data~\cite{Harris2019d,Siegle2021}---against conventional measures: lag-1 autocorrelation and standard deviation. %
We used experimental electrophysiological recordings of the mouse brain from the Allen Neuropixels Visual Behavior dataset~\cite{Allen2022}, which provides local field potential (LFP) data from ${\sim}20$ channels (between $15$ and $24$ for any given mouse) in each of six regions of the mouse visual cortex [shown schematically in \cref{fig:neuropixelsviz}]. %
We selected recording sessions based on quality metrics, extracted LFP data from periods where mice where viewing a static gray screen, applied a $1$--\SI{20}{Hz} bandpass filter, and finally decimated the signals to a sampling rate of \SI{125}{Hz}, yielding ${\sim}20$ time series (${\sim}5$\,min., or ${\sim}37,500$ samples long), for each of six visual cortical regions over 39 sessions (39 mice; see \cref{sec:neuropixelsupplement} for details).
We then calculated \newfeature{}, lag-1 autocorrelation, and standard deviation for each time series.
Note that we used centered \newfeature{}, \cref{eqn:centerednewfeature}, appropriate for these data with negative values.
To rank regions according by their position in the anatomical hierarchy, we used data on the hierarchical level from \citet{Siegle2021}, who recomputed the connectome-based scores from \citet{Harris2019d} for the visual regions of the Allen Neuropixels dataset [see \cref{fig:hierarchyviz}]. %
We performed a one-way analysis-of-variance on feature values (averaged across channels for a given probe) against cortical regions (using data pooled across all mice) to determine if there are significant differences in the distribution of feature values between regions.
We identified significant effects for each of centered \newfeature{} [$F(5, 190) = 66$, $p < 10^{-38}$], standard deviation [$F(5, 190) = 14$, $p < 10^{-11}$], and lag-1 autocorrelation [$F(5, 190) = 4$, $p = 1.5 \times 10^{-3}$], encouraging more detailed analysis on the variation of these features across the hierarchy.

We then sought to determine, at the level of individual mice, the strength of the relationship between the values of a given feature (computed across LFP channels) and the hierarchical ordering of the cortical areas of each channel.
We first calculated Kendall's $\tau$ (a tie-robust correlation coefficient) between feature values and hierarchical ranks across all channels, as shown for a single representative mouse in \cref{fig:corticalcorrelations_individual}. %
We used a permutation-based procedure to estimate a $p$-value for each computed $\tau$ statistic, generating an empirical null distribution for each feature by shuffling, for each mouse, the hierarchical ranks assigned to each region (ensuring that channels from the same region retain the same rank, thereby accounting for spatial correlation between nearby channels). %
After calculating a $\tau$ coefficient for each of $10^6$ independently shuffled datasets, $p$-values were estimated as the proportion of surrogate correlations, $\tau$ that were larger in magnitude than the $\tau$ statistic of the real data.
We found that regions lower in the visual hierarchy (e.g., the primary visual area, `VISp') have a lower value of centered \newfeature{}---corresponding to a greater DTC---than higher regions (e.g., the anteromedial visual area, `VISam'). %
To assess this relationship across mice, we performed a Mann-Whitney U test between the distributions of correlations for the original and shuffled datasets.
As shown in \cref{fig:corticalcorrelations_distribution}, neither standard deviation nor lag-1 autocorrelation have single-mouse correlation coefficients that are significantly different from their corresponding null distributions (at a threshold of $p = 0.05$).
By contrast, centered \newfeature{} achieved strikingly strong correlations for most mice (a median $\tau = 0.55$), and was highly significant ($p < 10^{-18}$).
That this signature is detectable by the noise-robust \newfeature{} but not lag-1 autocorrelation or standard deviation is consistent with the presence of variable noise levels across regions, masking the effects of criticality from traditional features.

Our results above provide evidence consistent with the hypothesis that, in a given mouse, brain regions higher in the visual hierarchy are closer to criticality.
We next aimed to test whether the result also holds across mice, i.e., that variability due to the hierarchical variation detectable by \newfeature{} is stronger than inter-mouse variability.
To achieve this, we pooled the data from all channels in all mice to calculate a group-level $\tau$ coefficient for each feature [shown in \cref{fig:corticalcorrelations_pooled}].
We estimated the statistical significance of group-level correlations using a similar permutation-based procedure as above (i.e., calculating the proportion of $10^{6}$ shuffles that have a correlation greater than the true measured value for a feature).
Remarkably, centered \newfeature{} remains highly correlated to the visual hierarchy even at the group-level ($\tau = 0.42$, $p < 10^{-6}$), whereas standard deviation ($\tau = -0.01$, $p=0.8$) and lag-1 autocorrelation ($\tau = -0.01$, $p=0.5$) did not detect any significant relationship, as per the individual-level analysis.
Our results suggest that higher-order cortical regions exhibit dynamics consistent with being closer to criticality~\cite{Hilgetag2020,Cocchi2017}, even though they may be influenced by varied levels of noise due to external input, cellular composition, laminarity, or the presence of vasculature. %
Together with the hypothesis that sensory regions are differentiated by their proximity to criticality, these results provide strong evidence that \newfeature{} can be a more reliable metric of the DTC than conventional metrics in noisy, real-world systems.

\section{Discussion}
\label{sec:Discussion}

This work addresses the challenge of estimating the distance to a critical point, DTC, in the presence of an uncertain, and potentially variable, confounding \noisestrength{}.
While many studies have tackled the challenge of describing how noise disrupts conventional metrics of the DTC~\cite{Meunier1988, Kuehn2011, Kuehn2013, ORegan2018} or intensifies low-amplitude perturbations by stochastic resonance~\cite{Lindner2004}, to our knowledge no work has attempted to identify new features that are insensitive to changes in the noise level.
Moreover, our work demonstrates the ability of our novel data-driven methodology to motivate new theory and algorithmic implementations relevant to working with real (finite and noisy) time series. %
Specifically, by comparing the performance of thousands of time-series analysis features, we identified analysis methods that were able to robustly track the DTC in systems with variable levels of dynamical noise, but which have not been applied to this problem in the past. %
By analyzing the algorithmic steps underlying top-performing time-series features, we developed a deeper theoretical understanding that enabled us to reverse-engineer a novel, efficient, and noise-robust index of the DTC that out-performs all other \textit{hctsa} features on this problem: the rescaled auto-density, \newfeature{}.

Our data-driven approach to solving theory-based problems is highly flexible, and could be extended to many real-world problems, particularly those involving short, noisy time-series data.
The approach involves first simulating a known dynamical mechanism (to generate time-series data with a known structure), and then searching across a sufficiently large and comprehensive set of candidate time-series features for those that can best recover the underlying structure.
This highly comparative methodology has been used previously by \citet{Fulcher2013} to find statistical estimators of the scaling exponent of self-affine time series, and the Lyapunov exponent of Logistic Map time series; problems for which high-performing features were known to exist within the candidate feature set.
This is also the case for the first fixed-noise setting investigated here, where our data-driven approach recapitulated the strong performance of conventional criticality metrics related to autocorrelation and the distribution of values.
However, here we also extended it to a new setting, the variable-noise setting, in which it was not known whether any features would exhibit strong performance.
We showed that conventional metrics with high \rhofix{} (which strongly correlate with the DTC in the fixed-noise setting) perform poorly in this variable-noise setting, with a low \rhovar{} (\cref{sec:variablenoise}).
But, surprisingly, we identified several high-\rhovar{} features that can track DTC in the presence of confounding variations in \noisestrength{}.

Having identified noise-robust features of DTC from \textit{hctsa}, we then showed how our data-driven methodology can motivate the development of new theory and understanding.
While a list of top-performing features is already a valuable resource for addressing the problem of inferring the DTC---and brings about the possibility of using an ensemble of metrics for a more accurate estimate---we also sought practical insight regarding these top-performing metrics, the problem itself, and the broader context of criticality in noisy systems. %
In this capacity, our approach sits alongside recent efforts to generate practical understanding from data using interpretable machine-learning algorithms~\cite{Gunning2019,Linardatos2020} or to solve analytical problems using artificial intelligence~\cite{Davies2021}.
Unlike existing approaches, our methodology uses a simple search across a comprehensive library of transparent algorithms drawn from existing literature, automatically flagging those that are relevant to a given problem.
Studying the highlighted algorithms can then yield new theoretical insight into the problem at hand.
In this work, noticing key algorithmic similarities in the top-performing features motivated us to develop a theoretical account of how these algorithmic steps were able to track DTC so successfully.
The resulting theory---invoking a potential formulation, Eq.~\eqref{eqn:SupercriticalHopfPotential}, and the corresponding expression for the invariant density, Eq.~\eqref{eqn:InvariantDensity}---was used to formulate a new high-performing time-series feature, \newfeature{}, that robustly infers the shape of the potential function, which depends only on the DTC, by measuring the time-series distribution after rescaling values with the spread of differences (see \cref{sec:NewFeature}).
\newfeature{} performs well in both noise settings, having a higher correlation to the DTC than all other features in the variable-noise setting, as well as a competitive performance (relative to conventional metrics) in the fixed-noise setting.
As such, \newfeature{} requires less prior knowledge of a system to produce an accurate estimate of the DTC, making it a practical statistic for tracking the DTC in real-world settings.
Given the increasingly broad and detailed datasets being generated across scientific domains, this work thus demonstrates an ability to derive theoretical insight and develop practical analytic tools via the broad algorithmic comparison enabled by large algorithmic libraries like \textit{hctsa} \cite{Fulcher2017}.
Our approach serves as a model for using wide methodological comparison to tackle similar problems that aim to develop new theory for bridging dynamical mechanisms with the statistical properties, such as criticality, that are most sensitive to the theoretical structures of interest.

No previous studies have tackled the challenge of developing noise-robust indicators of the DTC by distilling theory from data-driven exploration.
However, the algorithmic components of \newfeature{}, including the spread of differences and a measurement of asymmetry in the distribution, \cref{eqn:NewFeature}, share similarities with existing metrics related to criticality.
For instance, the spread of differences (see \cref{sec:Sigma}) has been used to anticipate critical transitions in cryptocurrency markets~\cite{Tu2022}, whereas skewness and related properties of the distribution (see \cref{sec:invariantdensity}) have been used to mark abrupt changes in ecosystems~\cite{Xie2019} and climates~\cite{Guttal2008}.
Crucially, neither the spread of differences nor the distribution alone give noise-robust features.
As we highlight here, the confounding influence of the \noisestrength{} can only be eliminated by carefully combining these two properties (see \cref{sec:NewFeature}).
Furthermore, estimating the DTC typically requires precisely calibrating a feature against the control parameter and the \noisestrength{} through repeated observation on a system-by-system basis.
Although \newfeature{} still requires calibration to the control parameter, unlike typical indicators of the DTC, there is less need to recalibrate for changes or uncertainty in noise (see \cref{fig:FeatureBubbles} for an illustration of how concrete predictions of the DTC can be made using linear fits of the top variable-noise \textit{hctsa} features).
Moreover, \newfeature{} is superior even in contexts that do not require concrete estimates of the DTC; we showed in \cref{sec:casestudy} that being robust to noise allows \newfeature{} to more accurately recover the relative positioning brain regions in the mouse visual hierarchy.
\newfeature{} also has many other advantages over other existing critical indicators that have been applied in noisy settings, including:
(i) it does not require perturbing the system~\cite{Lim2011};
(ii) it operates on univariate time series~\cite{Weinans2021}; and
(iii) it performs well on short time series (but assumes a small sampling period).
Many real-world systems are corrupted by noise with an unknown strength, can only be recorded for short periods, and cannot be measured in their full multivariate complexity.
As such, \newfeature{} may improve accuracy in the wide range of tasks that involve quantitatively anticipating criticality from time-series data.

We verified the real-world utility of \newfeature{} by applying it to a dataset of electrophysiological recordings from the mouse visual cortex, where it outperformed conventional metrics of the DTC in tracking the anatomical hierarchy of visual cortical regions (in \cref{sec:casestudy}). %
A long-standing hypothesis on cortical organization is that higher-order regions are closer to criticality, facilitating the integration of multiple external signals or stimulus features~\cite{Cocchi2017,Hilgetag2020,liang2024}, but this hypothesis has lacked concrete experimental evidence. %
Our finding that \newfeature{} increases along the mouse visual hierarchy offers a mechanistic explanation for the escalating gradient of slow fluctuations along the visual hierarchy as being caused by critical slowing down, with regions higher in the hierarchy exhibiting a smaller DTC.
Furthermore, our results suggest that in real-world neuronal systems, noise is better represented heterogeneously, with variable rather than fixed intensity.
Variable noise renders conventional time-series metrics of criticality, which are highly sensitive to variations in noise level between brain regions, blind to variation in the DTC. %
Future work could aim to extend our analysis of heterogeneity in the DTC to the broader cortex, incorporating LFP recordings from other cortical areas or whole-cortex data using imaging modalities with a wider field of view~\cite{Markicevic2021}. %
Moreover, Changes and disruptions to the DTC have been implicated in broader neural contexts~\cite{Zimmern2020}: for insomnia disorder and the transition to sleep~\cite{Yang2016}; autism spectrum disorder~\cite{Trakoshis2020a}; and epilepsy, at the transition to seizure~\cite{Maturana2020,Liu2023}. %
In these instances, employing \newfeature{} could also present a pertinent and innovative approach. %
For seizure anticipation especially, simple but reliable indicators that can be applied to noisy, non-invasive recordings are essential for building practical monitoring devices~\cite{Maturana2020,Liu2023}. %
Having been formulated for robustness against dynamical noise (in \cref{sec:NewFeature}), verified against moderate levels of measurement noise (in \cref{sec:robustnesssupplement}), and demonstrated on experimental electrophysiological data (in \cref{sec:casestudy}), \newfeature{} opens the door to studying criticality in noisy real-world settings and thereby connecting our measurements of the world around us to the deeper mechanistic principles that underlie them.

Even though our data-driven methodology for finding noise-robust features can be generically applied to various critical systems, we have made a number of simplifying choices that limit how well \newfeature{} will perform on arbitrary critical systems.
Foremost, we chose to examine a simple normal form [see \cref{sec:ModelSystem} and \cref{eqn:SupercriticalHopfRadial}] that describes a broad range of systems, from auditory hair cells~\cite{Ospeck2001,OMaoileidigh2012} to financial markets~\cite{Gao2009}, and many others~\cite{Marsden1976, Freyer2012}.
Features that perform well for the radial part of the supercritical Hopf bifurcation, \cref{eqn:SupercriticalHopfRadial}, can be adapted for the full form of the Hopf bifurcation by calculating the radius, and for the pitchfork bifurcation by taking the absolute value; we verified that this centering step allows \newfeature{} to translate to the pitchfork bifurcation in \cref{sec:robustnesssupplement}.
Although numerical simulations were only performed for normal forms, we expect our summary feature to perform well on other systems that exhibit a Hopf or pitchfork bifurcation.
However, other normal forms can exhibit fundamentally different changes during bifurcation than the Hopf or pitchfork normal forms.
The saddle-node normal form, for instance, occurs when an unstable and a stable equilibrium annihilate one another. %
In this case, the potential grows more asymmetric as the critical point is approached: unless the noise is sufficiently weak the DTC is large, \newfeature{} is unlikely to remain noise-robust for saddle-node bifurcations (and other systems with asymmetric normal forms, such as for transcritical bifurcations).
Nevertheless, many systems---such as the saddle-node, transcritical, and subcritical Hopf or pitchfork bifurcations~\cite{Kuehn2011}---exhibit critical transitions in which proximity to the critical point corresponds to explosive jumps toward distant attractors.
For these systems, time-series features are unable to give a deterministic estimate of the time to catastrophe at high values of $\eta$, since crossings of the unstable threshold can be induced by noise well before the critical point~\cite{Kuehn2011}.
The DTC is still a useful quantity, however, for inferring the likelihood of a critical transition, and given that most potential functions are locally quadratic around stable fixed points we expect \newfeature{} to generically out-perform conventional metrics under variable-noise conditions, in particular for subcritical Hopf and pitchfork bifurcations.
Applying our data-driven methodology to find the most noise-robust DTC indicators for new normal forms is a promising avenue for future work, along with using our approach across classes of bifurcations to find a critical indicator that is not only noise-robust, but remains consistent over classes of bifurcations~\cite{Grziwotz2023}.

In addition to a simple deterministic component, we have also considered a highly simplified noise process.
Additive dynamical noise, which here is Gaussian, independent, and identically distributed, appears in many systems and is a common modelling assumption.
Unlike measurement noise, which is incorporated into the signal after a system has evolved and been measured, dynamical noise is present in the equations of motion for a system, and continually interplays with the deterministic dynamics controlled by the potential function.
We verified that \newfeature{} is robust to both additive measurement noise, as well as low levels of colored dynamical noise, in \cref{sec:robustnesssupplement}. %
Nevertheless, critical systems can possess non-Gaussian noise~\cite{Fronzoni1987, Kaur2022}, multiplicative noise, or even noise in the control parameter itself~\cite{Freyer2012}; we did not study such cases here.
Moreover, we did not aim to find features that are insensitive to variation in both the \noisestrength{} and the sampling period, although we expect this to be a more difficult problem.
It is also crucial to recognize the difference between noise with an unknown or uncertain strength, as treated in this work, and noise with a fast, time-dependent strength, which would certainly interfere with all of our noise-robust features.
Regardless, we expect our results will generalize well to systems that: i) are near Hopf or pitchfork bifurcations; ii) are sufficiently close to a potential minimum, with no extreme perturbations; and iii) have a noise level small enough for the system to remain localised around a single potential minimum.
For systems outside of this regime, the methodology we used to develop our summary feature---performing a data-driven exploration of candidate features to uncover new theoretical principles---is viable for phase transitions and noise processes of other varieties, and is a promising avenue for future scientific work.

Given the inherent stochasticity and complexity of real-world critical systems, accurately tracking the DTC using statistical properties of time-series data poses a significant challenge, prompting interdisciplinary research efforts spanning several decades.
Most work has relied on two substantial assumptions: that the DTC is small enough for the theory of normal forms to apply, and that the dynamical noise is negligible or fixed.
However, conventional metrics of the DTC are highly sensitive to the more realistic setting of variable, or uncertain, noise.
In this work, we have used a powerful and thorough data-driven approach that surveys a vast library of time-series features to:
i) confirm that conventional metrics are disrupted by a variable \noisestrength{};
ii) uncover unstudied time-series features that are insensitive to the \noisestrength{};
iii) scrutinize these noise-robust features to develop new theoretical insight; and
iv) summarize our new understanding into a simple new feature for accurately inferring the DTC in noisy real-world systems.
This work thus demonstrates a pragmatic, data-driven way of understanding theoretical systems through simulated data and wide methodological comparison, which can automatically flag promising algorithms to motivate the development of new theory.
The result of this process, \newfeature{}, introduced here, is a viable measure of DTC for realistic settings of systems corrupted by unknown, and in general variable, \noisestrength{}.
We expect these innovations to enable new applications of dynamical systems thinking to noisy, real-world systems.

\subsection*{Code availability}

Matlab code for reproducing our simulations and analyses is available at the \href{https://doi.org/10.5281/zenodo.8185428}{\texttt{Criticality}} repository on GitHub~\cite{Harris2023}.
This Matlab repository includes a script to reproduce the key figures from this paper as well as functions for \newfeature{} (located in the file \verb|RAD.m|) and \verb|fitSupercriticalHopfRadius_1| (at \verb|potentialDistributions.m|).
\newfeature{} is available as the \verb|CR_RAD| function in \href{https://doi.org/10.5281/zenodo.3927083}{\textit{hctsa}} v1.08~\cite{Fulcher2023} and the \href{https://doi.org/10.5281/zenodo.10039292}{TimeseriesFeatures.jl} Julia package~\cite{Harris2023a}.
RAD is also included in the default set of \textit{hctsa} features as \verb|CR_RAD_1| (centered \newfeature{} using the standard deviation of lag-1 differences), \verb|CR_RAD_2| (with lag-2 differences) and \verb|CR_RAD_tau| (using differences at a lag equal to the first zero-crossing of the autocorrelation function).
For the electrophysiological application  presented in \cref{sec:casestudy}, we accessed all Neuropixels Visual Behavior data through open software published by the Allen Institute for Brain Science (the \href{https://github.com/AllenInstitute/AllenSDK}{\texttt{allensdk}}) using the Julia-language wrapper \href{https://github.com/brendanjohnharris/AllenNeuropixelsBase.jl/}{\texttt{AllenNeuropixelsBase.jl}}~\cite{Harris2024}. %
Julia scripts to reproduce our procedure for filtering sessions, accessing local field potential (LFP) data, and performing the analysis presented in the main text are available in the accompanying \href{https://doi.org/10.5281/zenodo.8185428}{\texttt{Criticality}} repository~\cite{Harris2023}. %

\subsection*{Acknowledgments}

We thank Dong-Ping Yang for helpful discussions during the early stages of this project.

\appendix

\section{Inspecting high-performing features}  \label{sec:TopFeaturesSupplement}

Here we outline of four \textit{hctsa} features that robustly track the DTC in the presence of uncertain noise (as identified in \cref{sec:variablenoise}), which we have studied in close to detail to uncover the two algorithmic principles vital for noise-robust inference of the DTC (see \cref{sec:UnderstandingTopFeatures}). %
As described in \cref{sec:variablenoise}, we manually selected four features from the clustering analysis depicted in \cref{fig:VariableNoisePanels}, aiming for features that were algorithmically dissimilar and belonged to different clusters. %
We then studied these four top features in turn to uncover the common algorithmic components that allow for noise-robust inference of the DTC. %

\paragraph{Change in autocorrelation after discarding extrema.}
The first feature we investigated is \verb|DN_RemovePoints_max_01_ac1diff|, which measures the change in the autocorrelation of a time series after discarding extreme values.
This feature performs well in both the fixed-noise (\rhofix{}$ = 0.94$) and variable-noise (\rhovar{}$= -0.88$) settings. 
Acting on an input time series $\mbf{x}$, this feature first discards the largest 10\% of positive values (while maintaining the temporal ordering of the remaining data points) to produce a thresholded time series $\mbf{x}^\prime$.
Next, it calculates the lag-1 autocorrelation $r_1$ for both the original and thresholded time series, outputting the difference of the resulting values, as
\begin{equation}
    f_{\textrm{DN}}(\mbf{x}) = r_1(\mbf{x}) - r_1(\mbf{x}^\prime)\,,\ \mbf{x}^\prime =\{x_i\, |\, x_i < P_{90}(\mbf{x})\}\,,
\end{equation}
where $P_{90}(\mbf{x})$ is the 90th percentile of $\mbf{x}$.
Provided $\Delta t$ is small compared to timescale of the deterministic dynamics of a system, it follows from the covariance of the sum of random variables that the lag-1 autocorrelation is approximated by a simple function of only the variance of a time series, $\sigma^2(\mbf{x})$, and the variance of differences, $\sigma^2(\Delta \mbf{x})$.
Described by these two quantities, the lag-1 autocorrelation is given by
\begin{equation}
\label{eqn:decomposed_ac}
r_{1} \approx \frac{2\sigma^2(\mbf{x}) - \sigma^2(\Delta \mbf{x})/2}{2\sigma^2(\mbf{x}) + \sigma^2(\Delta \mbf{x})/2}\,.
\end{equation}
Thus \verb|DN_RemovePoints_max_01_ac1diff| relies on: i) properties of the tailedness of the distribution, namely how $\sigma^2(\mbf{x})$ differs from $\sigma^2(\mbf{x}^\prime)$; which are ii) calibrated against the variance of differences, $\sigma^2(\Delta \mbf{x})$. Hence the algorithmic components of this feature motivate our close study of these two properties, the distribution and the variance of differences, in \cref{sec:UnderstandingTopFeatures}.

\paragraph{Probability of two consecutive high values.}
A second high-performing feature, in both the fixed-noise (\rhofix{}$= 0.93$) and variable-noise (\rhovar{}$= 0.88$) cases, is \verb|SB_MotifTwo_mean_uu|, which counts the proportion of consecutive pairs of time-series values that are both above the mean: 
\begin{equation}
    f_{\textrm{SB}} = P(x_t > \bar{x} \,\cap\, x_{t+1} > \bar{x})\,,
\end{equation}
where $\bar{x}$ is the mean of $\mbf{x}$.
Since $\Delta \mbf{x}$ depends predominantly on the \noisestrength{}, as described in \cref{sec:UnderstandingTopFeatures}, the probability of a point crossing the mean over an increment in time depends on: i) its distance from the mean; and ii) the variance of its diffusive motion, given by $\sigma^2(\Delta \mbf{x})$.
Hence by counting the proportion of values that remain above the mean, \verb|SB_MotifTwo_mean_uu| depends on both the properties we have identified as crucial for robustly inferring the DTC: i) the proportion of values that are above the mean at the initial time point (related to the distribution); and ii) the probability of crossing for points above the mean (related to the variance of differences).

\paragraph{Change in distribution after moving-average filter.}
Our third noise-robust estimator for the DTC is \verb|PP_Compare_rav2_kscn_olapint|, which has a high \rhofix{}$= 0.88$ and \rhovar{}$= -0.87$.
This feature measures the change in the probability density of time-series values (via an overlap integral to a best-fitting Gaussian distribution) after it has been smoothed by a single-period, two-sample moving average:
\begin{equation}
    f_{\textrm{PP}} = \frac{O[p(x; M[\mbf{x}])]}{O[p(x; \mbf{x})]}\,,
\end{equation}
where $p(x; \mbf{x})$ is a kernel-density estimate for the distribution of $\mbf{x}$, $M$ is a two-sample moving-average filter, and $O$ is an overlap integral given by
\begin{equation}
    O[p(x)] = \sigma(\mbf{x}) \int  p(x) \varphi(\bar{x}, \sigma(\mbf{x}))  dx\,,
\end{equation}
where $\varphi(\mu, \sigma)$ is the probability density of a Gaussian distribution with mean $\mu$ and standard deviation $\sigma$.
The main effect of a moving average filter on the distribution is an increase in the density at medial values, which is a result of peaks and troughs in the time series being truncated and depends on the distribution of values.
Moreover, the magnitude of the effect of truncating the most extreme values also depends on the mean size of the fastest-timescale fluctuations as well as the frequency of extreme peaks, which are both determined by the variance of differences.

\paragraph{Comparison of extrema in time-series windows.}
Finally, we outline \verb|ST_LocalExtrema_l100_meanrat|, which has \rhofix{}$= 0.91$ and \rhovar{}$= -0.90$.
This feature finds the average of maximum and minimum values computed across non-overlapping, $100$-sample windows of a standardized ($z$-scored) time series, then returns the ratio of these two values.
From another perspective, this feature is a proxy for the relative extent to which the invariant density expands on either side of the mean as $\eta$ increases. This can be seen by noting that the minimum time-series value in each window is close to zero, such that when the time-series is standardized, a value of $0$ becomes ${-\overline{x}}/{\sigma(\mbf{x})}$. Therefore, standardization reduces \verb|ST_LocalExtrema_l100_meanrat| to the ratio of the non-normalized average maxima and the mean time-series value:
\begin{equation}
    \label{eqn:extrema_feature}
    f_{\textrm{ST}} = \frac{\langle \max[(\mbf{x}_w - \bar{x})/\sigma(\mbf{x})]\rangle}{\langle |\min[(\mbf{x}_w - \bar{x})/\sigma(\mbf{x})]|\rangle} \approx \frac{\langle \max[\mbf{x}_w - \bar{x}]\rangle}{ \bar{x}}\,,
\end{equation}
where $\langle \max[\mbf{x}_w]\rangle$ is the mean value of the maximum in each 100-sample window, and $\min[\cdot]$ is the minimum value.
As illustrated in \cref{fig:TheorySchematic}, the mean of the distribution (related to the standard deviation) depends on both the DTC and the \noisestrength{}. The average maximum within a window, however, depends on the rate at which the system diffuses within the short, 100-sample windows. Hence this final feature, in a similar way to the other top \textit{hctsa} features we have detailed, also compares a distributional property to the variance of differences.

\section{Case study methodology}  \label{sec:neuropixelsupplement} %

In \cref{sec:casestudy}, we demonstrated how centered \newfeature{} applied to electrophysiological data tracks the functional hierarchy of the mouse visual cortex where conventional indicators of criticality and dynamic range fail. %
Here we provide detailed methods for our case study, including a description of the Allen Visual Behavior Neuropixels recordings, our pipeline for accessing and pre-processing the open dataset, as well as our procedure for evaluating RAD and other metrics on these data. %

The Allen Neuropixels Visual Behavior dataset~\cite{Allen2022} comprises electrophysiological recordings from the mouse visual cortex, acquired with Neuropixels probes~\cite{Jun2017}, while the mouse is presented with a series of visual stimuli~\cite{AllenInstitute2022}. %
A Neuropixels probe houses a checkerboard array of electrodes, spaced vertically by \SI{20}{\micro\metre} along a linear shank \SI{1}{cm} in length~\cite{Jun2017}. %
Inserted into the mouse brain, each probe records the electrical activity at hundreds of sites. %
For the Allen Neuropixels Visual Behavior dataset, six probes were inserted at the retinotopic centres of six areas of the visual cortex in each mouse~\cite{AllenInstitute2022}, and raw recordings were filtered into the local-field potential (\SI{<1000}{Hz}) and neuronal spikes ($0.3$--\SI{10}{kHz}). %
Each of $~75$ mice were presented with a series of visual stimuli, including a blank gray screen, gabor patches, full-field flashes, and an image-change detection task, across two recording sessions. %
The Neuropixels Visual Behavior dataset, having a high recording quality and a large number of probes that cover the majority of areas in the visual cortex, provides a valuable opportunity to study the distribution of critical dynamics across the visual hierarchy. %

We selected recording sessions that had location-tagged channels in all six of the target visual areas, and mice that displayed no abnormal histology or activity.
Session filtering resulted in 39 sessions from 39 mice. %
From these 39 sessions, we selected all LFP channels located in the cortex, giving between $15$ and $24$ channels for each visual area, for each mouse. %
We accessed LFP time series for each channel during the `\verb|spontaneous|' stimulus, during which the mouse was presented with a static gray screen. %
The resulting LFP time series were 5 minutes in length and sampled at a rate of \SI{1250}{kHz}. %
We then decimated the signal to a sampling rate of \SI{125}{Hz}, after bandpass filtering between $1$--\SI{20}{Hz} to: (i) avoid high-frequency down-sampling artifacts; as well as (ii) capture the dominant low-frequency activity in the theta ($4$--\SI{8}{Hz}) band alongside a portion of the higher-frequency $1/f$ activity. %
Finally, we computed centered \newfeature{}, \AC{}, and standard deviation for each cortical channel, for each session, before comparing to the structural hierarchy (as presented in~\cref{sec:casestudy}). %
An exact record of the filtering thresholds and the resultant sessions can be found in the \texttt{paper/Criticality.jl} folder of the accompanying code repository~\cite{Harris2023}, alongside scripts and data for reproducing~\cref{fig:casestudy}. %

\providecommand{\noopsort}[1]{}\providecommand{\singleletter}[1]{#1}%

\end{document}


\makeatletter

\makeatother 

\title{Tracking the distance to criticality in systems with unknown noise}
\author{Brendan Harris}
\affiliation{School of Physics, The University of Sydney, NSW 2006, Australia}
\author{Leonardo L. Gollo}%
\affiliation{%
The Turner Institute for Brain and Mental Health, School of Psychological Sciences, and Monash Biomedical Imaging, Monash University, Victoria 3168, Australia
}
\author{Ben D. Fulcher}
\affiliation{
 School of Physics, The University of Sydney, NSW 2006, Australia
}
\date{\today}

\onecolumngrid
\begin{center} \huge{\textsc{Supplemental material}}
\end{center}
\maketitle
\onecolumngrid
\vspace{-2em}
\addexternaldocument{main}

\section{Feature performance}
\label{sec:Supplement_results}

\begin{figure*}[htbp!]
    \centering\includegraphics[width=\textwidth]{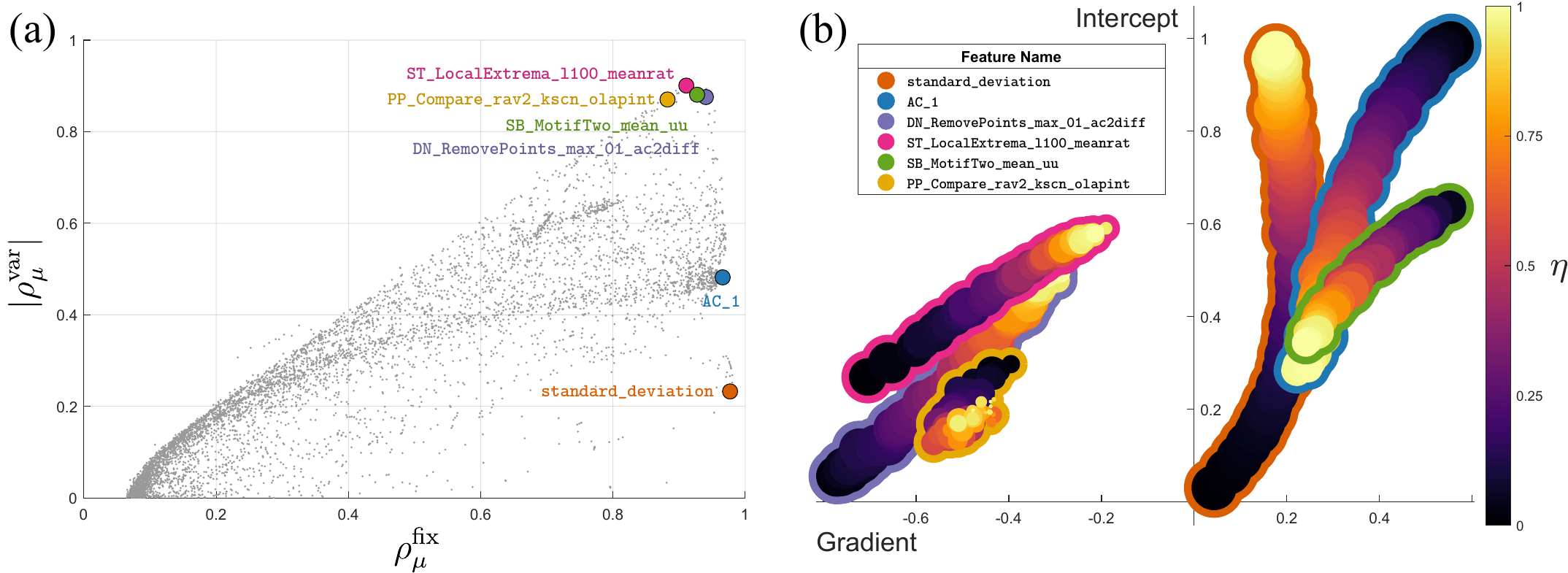}
    {\phantomsubcaption\label{fig:fixedvariableScatter}}
	{\phantomsubcaption\label{fig:FeatureBubbles}}
	\cprotect\caption{The top \textit{hctsa} features out-perform conventional metrics at producing concrete inferences of the DTC.
    (a) Each feature in the \textit{hctsa} library is plotted with a fixed-noise (\rhofix{}) and variable-noise (\rhovar{}) score.
    \verb|standard_deviation| (blue) has a large \rhofix{}$= 0.98$, indicating a strong correlation to the control parameter for fixed noise, but does not consistently predict $\mu$ when noise is not fixed (\rhovar{}$= 0.23$). The same is true for \verb|AC_1| (lag-1 autocorrelation, red), with \rhofix{}$= 0.97$ and \rhovar{}$= 0.48$.
    A small number of \textit{hctsa} features perform well in both settings, four of which are annotated.
    (b) Lines of best fit summarize the performance of the top \textit{hctsa} features as $\eta$ is varied.
    After rescaling values for each feature to the unit interval, least-squares linear regression was used to find their gradient ($x$-axis) and intercept ($y$-axis) against $\mu$ for each $\eta$.
    Each $\eta$ for each feature is represented by a bubble with a radius corresponding to a Spearman correlation against $\mu$ ($0.79 \le \abs{\rho_{\mu}} \le 0.99$). Features that perform well for fixed noise have large radii for all $\eta$ but may have a wide span.
    Note that at the scale $-1 \le \mu \le 0$, \rhovar{} is more sensitive to the intercept than the gradient.
    }
\end{figure*}

Here we present additional results that further characterize the performance of the \textit{hctsa} features in the fixed-noise (from \cref{sec:FixedNoise}) and variable-noise (from \cref{sec:variablenoise}) cases.
In \cref{fig:FixedNoiseFeatureHistogram} we showed how the \textit{hctsa} features are distributed over fixed-noise scores, and \cref{fig:VariableNoiseFeatureHistogram} presented the distribution of variable-noise scores.
To extend these findings here, in \cref{fig:fixedvariableScatter} we show  how the \textit{hctsa} features are distributed jointly over the fixed-noise, \rhofix{}, and variable-noise, \rhovar{}, scores.
All \textit{hctsa} features reside in the lower right region of \cref{fig:fixedvariableScatter}, confirming that no features have a high variable-noise score, \rhovar{}, but a low fixed-noise score, \rhofix{}.
Since all of the top variable-noise features also perform strongly in the fixed-noise case, this result motivates our decision to choose the top features based on the variable-noise score, \rhovar{}.

Although \rhovar{} provides a simple way to compare features on their robustness to uncertain \noisestrength{}, it is important to note that this statistic is only a summary of Spearman correlations between $\mu$ and the values of each feature for a chosen range of control and noise parameters.
As such, while \rhovar{} scores features on how well they monotonically track the DTC, it does not directly capture the more challenging process of inferring the DTC directly from data.
Such challenges include heteroskedasticity in the relationship between feature values and $\mu$, the optimal model for predicting the DTC for a given feature, and how such a model generalizes to other parameter ranges.
Given these challenges, we aimed to better understand how the ability of the top features to infer the DTC varies across \noisestrength{}s.
To estimate the DTC from a time series, we computed a linear fit (comprising a gradient and an intercept) between each feature's value and the control parameter $\mu$, for each \noisestrength{}.
We then visualized the behaviour of each feature for various values of $\eta$ as a trajectory in gradient--intercept space, shown in \cref{fig:FeatureBubbles} for the six \textit{hctsa} features that we investigated in detail in the main text.
We found that the top-performing \textit{hctsa} features are distinct from the two conventional estimators of criticality, standard deviation and lag-1 autocorrelation.
In \cref{fig:FeatureBubbles}, features that perform well in the variable noise setting span a smaller region of the gradient--intercept space, which can counteract lower fixed-noise correlations.
For example, \verb|PP_Compare_rav2_kscn_olapint| is noise-robust since its relationship to $\mu$ is highly consistent, despite a smaller correlation for each $\eta$, while \verb|SB_MotifTwo_mean_uu| spans a larger region but has greater fixed-noise correlations.

\Cref{fig:FeatureBubbles} also allows us to make further qualitative judgements about the top features.
For instance, \verb|SB_MotifTwo_mean_uu| has a smaller span than \verb|DN_RemovePoints_max_01_ac1diff| but also a slightly smaller average correlation (see \cref{sec:UnderstandingTopFeatures}). This observation indicates that \verb|SB_MotifTwo_mean_uu| is a more suitable feature than \verb|DN_RemovePoints_max_01_ac1diff| when the uncertainty in $\eta$ is large.
Moreover, \verb|DN_RemovePoints_max_01_ac1diff| and \verb|SB_MotifTwo_mean_uu| exhibit reliably strong correlation to $\mu$ across our chosen range of $\eta$, whereas \verb|ST_LocalExtrema_l100_meanrat| and \verb|PP_Compare_rav2_kscn_olapint| perform optimally at low $\eta$ but degrade as $\eta$ increases.
This trade-off between performance in specific and general scenarios suggests a degree of theoretical dissimilarity between our top features, which could be explored in future work to engineer indicators that are specialised to specific noise levels, or ensembles of simple features that together have superior performance across a range of noise strengths.
Here, however, we have focused on the similarities between these features with the aim of developing a generically noise-robust feature.

\vspace{2em}

\newpage

\section{RAD robustness}  \label{sec:robustnesssupplement} %

In \newfeature{}, we have developed a novel noise-robust metric for tracking criticality through a data-driven approach applied to a simple, but largely generic normal form. %
As discussed in \cref{sec:Discussion}, we chose to study the radial component for the normal form of a Hopf bifurcation since it readily generalizes to the pitchfork normal form and exhibits no telling explosive behaviors. %
We also chose a specific form of dynamical noise: white, i.i.d. noise with a Gaussian distribution, which simplifies our theoretical findings and is the most common form of noise in both theoretical and experimental literature. %
In this section, we demonstrate that \newfeature{} out-performs conventional indicators of criticality even in other systems and with different forms of noise. %
We first compare \newfeature{}, lag-1 autocorrelation, and standard deviation on an alternative normal form, for a pitchfork bifurcation, before testing other forms of dynamical noise: pink colored noise and Brownian noise. %
Finally, we introduce additive, white, Gaussian measurement noise to our radial Hopf system and study how each metric of the DTC performs at different strengths of measurement noise. %
\newfeature{} performs well in most cases, but we note that in any scenarios where the lower bound and mode of the distribution are not equal, \newfeature{} requires a centering step: subtracting the median of the time series, then taking the absolute value [as centered \newfeature{}, \cref{eqn:centerednewfeature}]. %

\paragraph{Quadratic potential}

\begin{figure*}[h]
    \centering\includegraphics[width=0.9\textwidth]{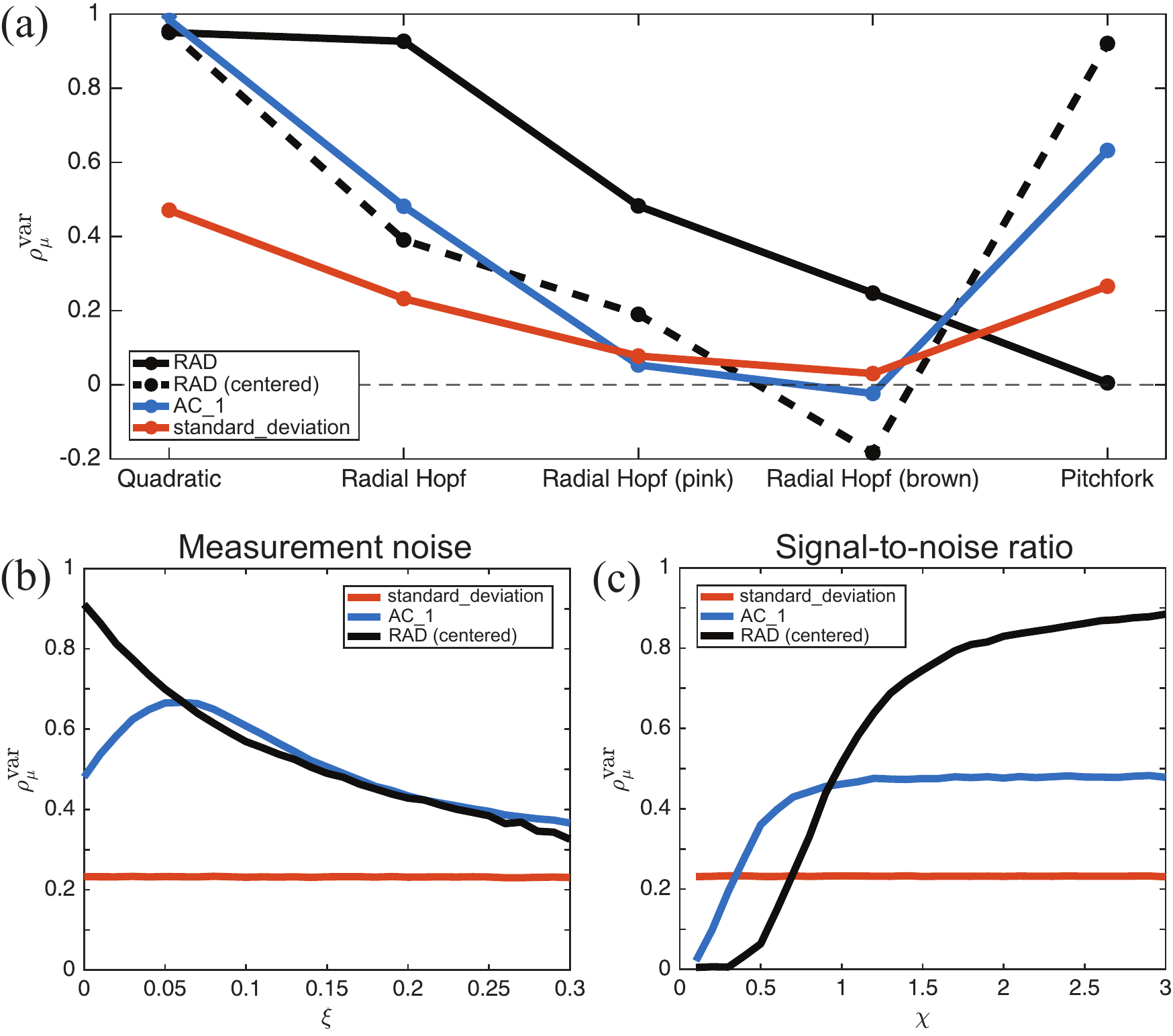}
	{\phantomsubcaption\label{fig:othersystems}}
    {\phantomsubcaption\label{fig:measurementnoise}}
	{\phantomsubcaption\label{fig:signaltonoise}}

	\cprotect\caption{\newfeature{} achieves higher variable-noise scores, \rhovar{}, than conventional critical indicators across different systems, forms of dynamical noise, and strengths of measurement noise.
    (a) The variable-noise score \rhovar is plotted for \newfeature{}, a centered version of \newfeature{}, standard deviation, and lag-1 autocorrelation (\AC{}), across a system with a quadratic potential, the radial Hopf system with pink and Brownian dynamical noise, and the normal form for a pitchfork bifurcation (see \cref{sec:robustnesssupplement}). %
    \newfeature{} performs well on radial systems, but requires a centering step for the symmetric pitchfork normal form.
    (b) \rhovar{} for standard deviation, \AC, and \newfeature{} (centered) as a function of the amplitude $\xi$ of white, Gaussian measurement noise added to the radial Hopf system. %
    (c) \rhovar{}, for the the same metrics of the DTC, as a function of the signal-to-noise ratio $\chi$ of the radial Hopf system with measurement noise. %
    }
\end{figure*}

We first verify the performance of \newfeature{} on a simple system with a quadratic potential, given by the Ornstein--Uhlenbeck process
\begin{equation}
    dx = \mu x\, dt + \eta\, dW\,,\quad x \ge 0\,. %
\end{equation}
For this system, the invariant density is Gaussian and the autocorrelation is independent of $\eta$, meaning \newfeature{} should become redundant with lag-1 autocorrelation, \AC{} (cf. \cref{sec:NewFeature}, and our use of the scale-invariance of the dynamics under a quadratic potential to derive \newfeature{}).
As shown in \cref{fig:othersystems}, \newfeature{} and lag-1 autocorrelation have comparable variable-noise scores for the quadratic potential. %

\paragraph{Supercritical pitchfork bifurcation}

The supercritical pitchfork bifurcation is a common co-dimension one bifurcation; like the Hopf bifurcation, and unlike most other codimension-one bifurcations, it has no unstable branch prior to the critical point.
The supercritical pitchfork bifurcation has a normal form given by
\begin{equation}
    dx = \left(\mu x - x^3\right) dt + \eta\, dW\,. %
\end{equation}
This normal form is almost identical to the radial component of the normal form for the supercritical Hopf bifurcation, only differing in the possibility of taking negative values; the pitchfork normal form has a similar potential function to that of the radial Hopf bifurcation, without a reflecting boundary at the origin.
To account for the symmetric nature of the potential function for the pitchfork bifurcation, we modified \newfeature{} to include a centering step, in which we subtract the median of the time series before taking the absolute value [given by ~\cref{eqn:centerednewfeature} in \cref{sec:NewFeature}]. %
The resulting time series has a distribution that is the positive half of a symmetric, unimodal distribution, matching the conditions of our original formulation \newfeature{}.
\cref{fig:othersystems} shows that the uncentered \newfeature{} fails for the pitchfork normal form, whereas the centered version out-performs both \AC{} and standard deviation. %

\paragraph{Colored dynamical noise}
Many systems in nature are subject to colored noise, characterized by a power spectrum that decays as a power law with frequency. %
Of the forms of colored noise, pink noise---having a $1/f$ spectral scaling---is particularly common in biological systems~\cite{Szendro2001}, characterizing a driving force that is non-Markovian~\cite{KlosekDygas1988}. %
To test the performance of \newfeature{} on a system with $1/f$ noise, we substituted white noise in \cref{eqn:SupercriticalHopfRadial} with pink noise, giving %
\begin{equation} %
\label{eqn:SupercriticalHopfRadialPink} %
\frac{dx}{dt} = \left(\mu x - x^3\right) + \eta P(t)\,,\quad x \ge 0\,, %
\end{equation} %
where $P(t)$ is a quasi-Gaussian pink noise process, scaled to have unit variance such that $\eta$ gives the standard deviation of the driving noise. %
We then simulated \cref{eqn:SupercriticalHopfRadialPink} for $\mu = -1, -0.99, \dots, 0$, and $\eta = 0.005, 0.01, \dots, 0.5$, using the same procedure and integration parameters detailed in \cref{sec:TimeSeries}. %
Finally, we performed simulations with the same parameters, but using Brownian noise in place of pink noise:
\begin{equation} %
    \label{eqn:SupercriticalHopfRadialBrownian} %
    \frac{dx}{dt} = \left(\mu x - x^3\right) + \eta B(t)\,,\quad x \ge 0\,, %
\end{equation} %
where $B(t)$ is a unit-variance quasi-Gaussian Brownian noise process, having a power spectrum proportional to $1/f^2$. %
\cref{fig:othersystems} shows that all three selected metrics perform poorly for pink and Brownian noise, but in both cases \newfeature{} out-performs lag-1 autocorrelation (\AC{}) and standard deviation. %

\paragraph{Measurement noise}

In addition to dynamical noise that drives a system, experimental data are typically subject to measurement noise, which is independent of the system dynamics. %
To verify that \newfeature{} out-performs conventional indicators of criticality even in the presence of measurement noise, we added additive, white, Gaussian noise to time series generated from \cref{eqn:SupercriticalHopfRadial}. %
We tested against two forms of measurement noise: (i) noise with a fixed amplitude, given by the parameter $\xi$; and (ii) noise with signal-to-noise ratio given by $\chi$, where the signal-to-noise ratio is defined as the standard deviation of the original time series divided by the standard deviation of the measurement noise.
Since approaching criticality causes changes in the amplitude of the original time series, testing both forms of measurement noise covers experimental cases in which the measurement noise strength is either absolute or scales to the signal strength. %
In both cases of measurement noise, we simulated independent datasets of time series using \cref{eqn:SupercriticalHopfRadial} for $\mu = -1, -0.99, \dots, 0$, and $\eta = 0.01, 0.02, \dots, 1$. %
We then added measurement noise to each time series, using $\xi = 0.01, 0.02, \dots, 0.3$ for the case of fixed-strength measurement noise, and $\chi = 0.1, 0.2, \dots, 3$ for the signal-to-noise-ratio case. %
Finally, we computed \newfeature{}, standard deviation, and lag-1 autocorrelation across all time series. %
We used the centered version of \newfeature{},~\cref{eqn:centerednewfeature}, as for the pitchfork normal form and the case study given in \cref{sec:casestudy}, since measurement noise destroys the positive-only distribution of the original time series. %

The variable-noise performances of standard deviation, lag-1 autocorrelation, and centered \newfeature{} are plotted across the strength of measurement noise $\xi$ in \cref{fig:measurementnoise}, and across the signal-to-noise ratio $\chi$ in \cref{fig:signaltonoise}. %
We note that increasing $\chi$ corresponds to a greater proportion of true signal (the dynamics of the underlying stochastic system) to independent measurement noise, raising the performance of \newfeature{}, whereas increasing $\xi$ corresponds---on average---to a lower signal-to-noise ratio, lowering performance.
Nevertheless, centered \newfeature{} has a high \rhovar{} at most levels of measurement noise, matching or out-performing lag-1 autocorrelation for $\xi < 0.3$ (or $\chi > 1$).
Since the variance of two independent random processes add linearly, the performance of standard deviation is unaffected by stronger measurement noise in both cases.
The performance of lag-1 autocorrelation increases consistently with $\chi$ but, curiously, peaks at $\xi \approx 0.05$ before steadily reducing.
This peak results from two effects that compete, under our \rhovar{} performance score, at different measurement noise strengths. %
First, for a measurement noise strength close to zero, increasing the dynamical noise tends to decrease the autocorrelation (as for the noise-free case presented in \cref{sec:variablenoise}). %
Second, for strong measurement noise, increasing dynamical noise boosts the signal-to-measurement-noise ratio and thus increases the autocorrelation (as, unlike the measurement noise, the signal has non-zero autocorrelation).
For $\xi \approx 0.05$, these effects balance each other to give a minimal variation in autocorrelation across values of the dynamical \noisestrength{}.
In contrast, controlling the signal-to-noise ratio as $\chi$ [\cref{fig:signaltonoise}] eliminates the second effect, resulting in a monotonic variation of the performance of lag-1 aucorrelation. %
Nevertheless, centered \newfeature{} performs best across most values of measurement noise strength and signal-to-noise ratio. %

\providecommand{\noopsort}[1]{}\providecommand{\singleletter}[1]{#1}%